\documentclass[10pt]{elsart}
\makeatletter
\renewcommand{\@maketablecaption}[2]{%
\@tablecaptionsize
\global \@minipagefalse
\hbox to \hsize{\parbox[t]{\hsize}{#1 \\[1ex] #2\\[1ex]}}%
}%
\def\section{\@startsection{section}{1}{\z@}{0.2\@bls
  \@plus .4\@bls \@minus .1\@bls}{0.1\@bls}{\normalsize\bfseries}}
\def\subsection{\@startsection{subsection}{2}{\z@}{0.2\@bls
  \@plus .1\@bls \@minus .1\@bls}{0.1\@bls}{\normalsize\itshape}}
\def\subsubsection{\@startsection{subsubsection}{3}{\z@}{0.2\@bls
  \@plus .1\@bls}{0.0001pt}{\normalsize\itshape}}
\def\paragraph{\@startsection{paragraph}{4}{\z@}{1.25ex \@plus
  0ex \@minus 0.2ex}{-1em}{\normalsize\bfseries}}
\makeatother
\renewcommand{\appendix}{%
  \renewcommand{\section}{%
    \newpage\thispagestyle{plain}%
    \secdef\Appendix\Appendix}%
  \setcounter{section}{0}%
  \renewcommand{\thesection}{\Alph{section}%
  }
}
\newcommand{\Appendix}[2][?]{%
  \refstepcounter{section}%
  \addcontentsline{toc}{appendix}%
    {\protect\numberline{\appendixname~\thesection} #1}%
  {\bfseries\appendixname\ \thesection%
   \par}%
  \sectionmark{#1}
}
\usepackage{graphicx}
\usepackage[footnotesize]{subfigure}

\usepackage{chicago}
\usepackage{pifont}
\usepackage{textcomp}
\DeclareMathAlphabet{\mathpzc}{OT1}{pzc}{m}{it}
\usepackage{array}
\newcommand{\BIBand}{and}
\usepackage{amsmath}
\usepackage{amssymb}
\newcommand{\Gammap}  {\gamma\,'}
\newcommand{\Gammapp}  {\gamma\,''}

\begin{document}
\fontsize{11}{13.8}\selectfont
\begin{frontmatter}
\title{\large Are granular materials simple?
An experimental study\\ of
strain gradient effects and localization}
\author{Matthew R. Kuhn}
  \ead{kuhn@up.edu}
\address{Dept.\ of Civil and Env.\ Engrg., 
  Univ.\ of Portland,
  5000 N. Willamette Blvd., Portland, OR, U.S.A. 97203\\
  Tel.: 503-943-7361; Fax: 503-943-7316}
%
\begin{abstract}
Experiments test the dependence of shearing stress on the first two
gradients of shear strain.
The tests were conducted by direct numerical simulation using
the Discrete Element Method (DEM) on a large two-dimensional
(2D) assembly of circular disks.
The assembly was coerced into non-uniformly
deformed shapes by applying body forces to the material region.
The tests show that shearing stress is
affected by both the first and second gradients of shear strain,
and the measured responses to strain
and its gradients are all incrementally nonlinear.
The dilation rate is unaffected by strain gradients.
Particle rotations, although highly erratic, are, on average,
consistent with the mean-field rotation and unaffected by
strain gradients.
In independent unconstrained tests, the material was sheared without
body forces so that localization could freely occur.
Three localization patterns were observed:
microbands at very small strains;
non-persistent shear bands at moderate strains; and persistent
bands at large strains.
The observed features of microbands and shear bands
are consistent with the measured influences of 
shearing strain and its first two derivatives.
\end{abstract}
\end{frontmatter}
%
\section{Introduction} \label{sec:intro}
\citeN{Noll:1958a} defined a simple material as one in which the stress
at a material point depends upon the first
deformation gradient and its history at that point.
The sole dependence upon local strain as the kinematic variable
distinguishes simple materials from other, more generalized descriptions,
such as those in micro-polar, non-local,
and strain gradient models.
This paper describes experiments that directly demonstrate that
granular materials are non-simple and that shearing stress depends 
upon the first three gradients of
shearing displacement (i.e., the shearing strain and its first two gradients).
The experiments were conducted by coercing non-uniform deformations
and measuring the local material response.
In the paper, we also address the manner in which shearing deformation
can freely localize within a granular material and the relationship between
strain gradients and the localization patterns.
\par
\emph{Strain gradient effects} have been
postulated for granular materials, 
and many gradient-dependent constitutive formulations have already
been proposed.
Such generalized constitutive proposals have certain 
attractive features:
strain gradient models can be used
to construct continuum
descriptions of materials that are, in fact, discrete at a micro-scale;
non-simple continuum descriptions introduce 
a characteristic length that can be used to
describe scale-dependent localization phenomena; and
including higher order deformation gradients within a constitutive
framework can circumvent problems of ill-posedness 
in the associated boundary value problems.
Although gradient-dependent models offer these features,
there has been no direct experimental evidence that
granular materials exhibit such behavior.
Indirect experimental evidence has been provided with 
other materials, such as micro-indentation
testing and thin-wire torsional testing
(e.g., \citeNP{Fleck:1997a}).
With granular materials, however, we have the ability---perhaps unique
among materials---to observe, measure, and, hopefully, resolve this
complexity with numerical experiments that realistically simulate
the underlying micro-scale mechanics.
The primary means of performing these experiments is the Discrete Element
Method (DEM), which has become
an effective tool for directly simulating and exploring the
mechanical behavior of granular materials at both the particle and
macro levels \cite{Cundall:1979a}.
\par
The experiments in this paper reveal quite unusual and complex
behavior, behavior that has not yet been entirely captured
in any continuum description.
In one set of experiments,
we observe the localization patterns that spontaneously appear
during shearing.
In other experiments,
the effects of the first and second strain gradients
are measured by
coercing a large assembly of particles into
predetermined shapes.
The coerced shapes were achieved by applying body forces in a controlled
manner. 
The material response was
computed with averaging techniques that allow us to
characterize both the average response and the variability
of the response under conditions of non-uniform deformation.
The paper proceeds as follows:
\begin{itemize}
\item
Section~\ref{sec:methods} describes the granular assembly
and summarizes the experimental methods.
\item
Sections~\ref{sec:results}--\ref{sec:results_IV}
give the experimental results.
Section~\ref{sec:results} catalogs the localization patterns
that spontaneously appear during unconstrained
shearing.
Sections~\ref{sec:results_III} and~\ref{sec:results_IV}
report the cumulative and incremental 
gradient-dependent responses that were measured in constrained
non-uniform shearing.
\item
In Section~\ref{sec:gradients_localization}
we consider some characteristics of the localization
patterns that are reported in Section~\ref{sec:results}
and whether these characteristics are consistent
with the gradient-dependent behavior that is reported
in Sections~\ref{sec:results_III} and~\ref{sec:results_IV}.
\item
Section~\ref{sec:conclusions} summarizes the experimental
findings.
\end{itemize}
\section{Methods} \label{sec:methods}
\begin{figure*}
\centering\small
\includegraphics{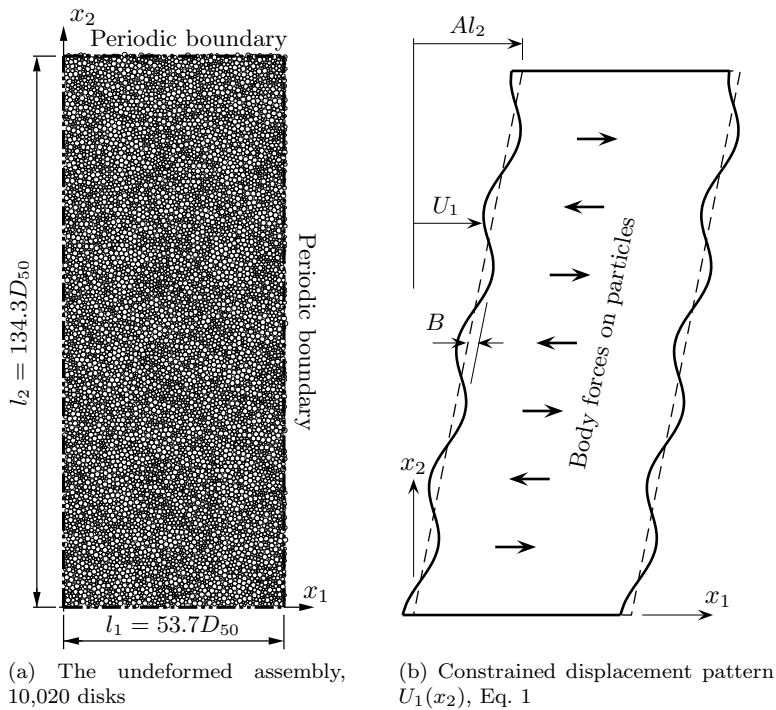}
\caption{The assembly of 10,020 disks
and a coerced horizontal displacement
profile, as in the Series~II--IV constrained shear tests.}
\label{fig:assemblies}
\end{figure*}

The work is primarily experimental.
Quasi-static numerical simulations
were performed on a large rectangular
two-dimensional assembly of 10,020 densely packed
circular disks (Fig.~\ref{fig:assemblies}a).
Two independent sets of shearing experiments were conducted.
The first set was conducted under \emph{unconstrained} conditions
in which localization could spontaneously occur. These are the Series~I
tests described in
Sections~\ref{sec:series_I_methods} and~\ref{sec:ushear}.
The second set of experiments was conducted under \emph{constrained}
conditions that permit the direct measurement of strain gradient effects.
These are the Series~II--IV tests of 
Sections~\ref{sec:series_II_method}--\ref{sec:series_IV_method}
and~\ref{sec:compare_con_uncon}--\ref{sec:dgamma_large}.
\fontsize{11}{13.8}\selectfont
\subsection{Description of the material}\label{sec:material_description}
The assembly height $l_{2}$ was 
about 134 particle diameters and about $2.50$
times its width $l_{1}$ (Fig.~\ref{fig:assemblies}a).
The particle assembly was surrounded on all sides by periodic boundaries.
The diameters of the 10,200 circular disks
were randomly distributed within a fairly narrow
range of $0.45$--$1.40$ times the median particle diameter $D_{50}$.
The particles were initially assembled and compacted 
into an irregular but macroscopically isotropic arrangement.
The initial assembly was fairly dense, with a void ratio of 0.172
and an average coordination number of 4.04.
The contact indentations were small~--- on average, about 0.11\% of
$D_{50}$.
\par
A simple force mechanism was employed at the particle contacts.
Linear normal and tangential contact springs were
assigned equal stiffnesses, and
the coefficient of contact friction was 0.50.
Unlike the model of \citeN{Iwashita:1998a},
no rolling resistance was included in the contact mechanism.
\fontsize{11}{13.8}\selectfont
\subsection{Constraining the deformations} \label{sec:methods_defs}
Both constrained and unconstrained shearing tests were conducted.
To study the effects of strain gradients, non-uniform shearing
deformations were coerced in a controlled manner by applying horizontal body
forces to particles throughout the assembly.
The numerical algorithm, a modification of the 
Discrete Element Method (DEM), produces a progressively
more deformed sinusoid shape (Fig.~\ref{fig:assemblies}b).
Individual particles, however, were allowed to freely
move and rotate
within their neighborhoods of particles, 
provided that, on average, the
particles moved in collective accord with the prescribed 
sinusoid contour.
The numerical algorithms are described elsewhere \cite{Kuhn:2003b}.
\par
In tests with constrained deformations,
the assembly was slowly and progressively sheared through a
sequence of
horizontal displaced shapes $U_{1}$ (Fig.~\ref{fig:assemblies}b),
\begin{equation} \label{eq:U1}
U_{1}(x_{2},\,t) = A(t)x_{2} + B(t) \cos
\left(
\frac{2\pi \,n}{l_{2}} x_{2} - \phi
\right)
\end{equation}
which were predetermined by the controlling shape 
parameters $A(t)$, $B(t)$, $n$,
and~$\phi$.
The horizontal shearing strain~$\gamma$ and its first two
gradients, $\Gammap$ and~$\Gammapp$,
are directly computed as
\begin{align}\label{eq:gamma}
\gamma(x_{2},t) &\equiv
\partial U_{1} / \partial x_{2} \\
&= A(t)-B(t) \frac{2\pi \,n}{l_{2}} \sin
\left(
\frac{2\pi \,n}{l_{2}} x_{2} - \phi
\right) \notag\\
\label{eq:gammap}
\Gammap(x_{2},t) &\equiv
\partial^{2} U_{1} / \partial x_{2}^{2} \\
&=
-B(t)  \left(\frac{2\pi \,n}{l_{2}}\right)^{2} \cos
\left(
\frac{2\pi \,n}{l_{2}} x_{2} - \phi
\right) \notag\\
%
\label{eq:gammapp}
\Gammapp(x_{2},t) &\equiv
\partial^{3} U_{1}/\partial x_{2}^{3} \\
&=
B(t) \left(\frac{2\pi \,n}{l_{2}}\right)^{3} \sin
\left(
\frac{2\pi \,n}{l_{2}} x_{2} - \phi \notag
\right) .
\end{align}
In each simulation experiment, 
the values of $n$ and~$\phi$ were held constant,
while $A$ and~$B$ were advanced in small steps.
The separate effects of $\gamma$, $\Gammap$, and~$\Gammapp$
were investigated by running numerous simulations with
different $n$ and $\phi$ values and by
slowly advancing the progressions of $A(t)$ and~$B(t)$
(Sections~\ref{sec:series_II_method}--\ref{sec:series_IV_method}).
\par
\begin{figure}
\centering
\includegraphics{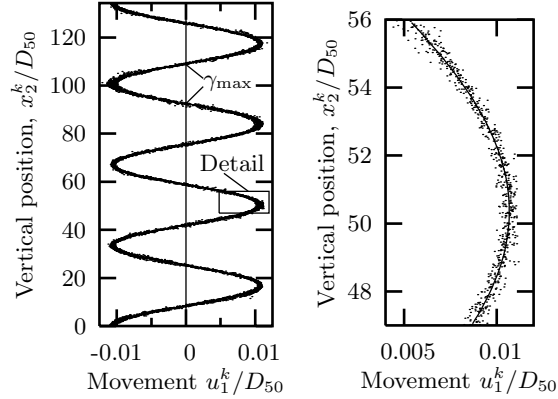}
\caption{Horizontal movements of the 10,020 
particles in a single Series~III sinusoid
shear test (Eq.~\ref{eq:U1}:  $n=4$, $\phi=0$, $A=0$, $B=-0.011D_{50}$)
Each dot represents the displaced center
of a single ($k^{\mathrm{th}}$) particle.}
\label{fig:movements_k}
\end{figure}

The assembly width~$l_{1}$ was maintained constant 
during all simulations,
but vertical dilation was freely allowed
under the condition of a constant vertical stress $\overline{\sigma}_{22}$
(Section~\ref{sec:dilatancy}).
The horizontal displacement $U_{1}$ in equation~(\ref{eq:U1}) 
represents a circumstance
that was only satisfied \emph{in the mean} by the 10,020 particles.
The movement of each, $k^{\text{th}}$ individual particle, 
$u_{1}^{k}$, was free to fluctuate about the
mean $U_{1}$, as is shown in Fig.~\ref{fig:movements_k}.
The particle movements in the figure
are measured relative to their 
positions in the undeformed assembly (at strain $\gamma = 0$)
and have been normalized by dividing by the median particle diameter
$D_{50}$, which we adopt as the fundamental unit of length.
%
\fontsize{11}{13.8}\selectfont
\subsection{Body forces and stress}%
\begin{figure}
\centering\small
\includegraphics{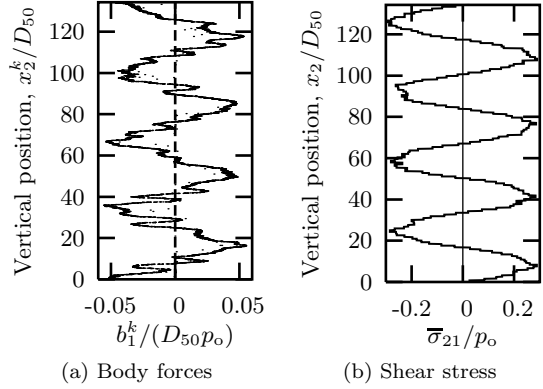}
\caption{Horizontal body forces and shear stress within the assembly for the 
displacement profile shown in Fig.~\ref{fig:movements_k}.}
\label{fig:bodyforce_stress}
\end{figure}

\label{sec:body_force_stress}
The constrained displacement contours~(\ref{eq:U1}) were produced
by applying a horizontal body force 
to each particle in the assembly \citeN{Kuhn:2003b}.
The algorithm produces an erratic distribution of body forces
$b_{1}^{k}$,
as is apparent in the 
scatter plot of Fig.~\ref{fig:bodyforce_stress}a.
We are primarily interested in the
horizontal shearing stress $\overline{\sigma}_{21}$ 
and its variation along the height of the assembly.
Because stress is highly non-uniform within granular materials,
averaging techniques
were required to arrive at a representative shearing stress 
at any vertical level $x_{2}$ within the assembly.
The horizontal shearing stress $\overline{\sigma}_{21}$
across a level $x_{2}$ was computed in the manner suggested
by \citeN{Bagi:1999a}.
This shearing stress
is an average \emph{local} stress,
computed across surfaces (or within thin horizontal regions) that encompass
about 80 particles,
which could be considered the size of the representative
volume element (RVE) of this study.
The shear stress within such small regions will likely
vary greatly within the assembly, and this variability is the topic of
Section~\ref{sec:variation}.
The rather substantial variability was partially moderated, however, 
with averaging
techniques that are described in Section~\ref{sec:series_III_method}.
\fontsize{11}{13.8}\selectfont
\subsection{Phase-space representations}\label{sec:space1}
\begin{figure}
\centering
\includegraphics{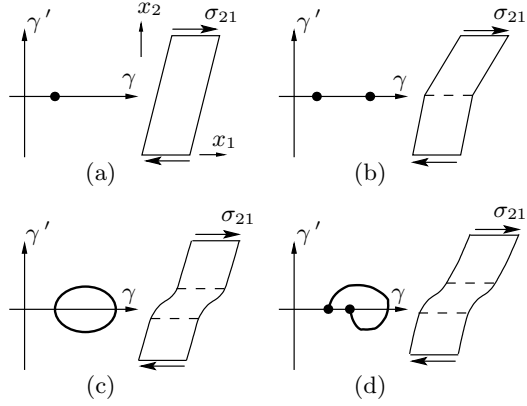}%
\caption{Phase-plane representations of 
shearing patterns.}
\label{fig:phase_spatial}
\end{figure}

A phase-space can be used to represent combinations of
shearing deformation $\gamma$ and its gradients
$\Gammap$ and $\Gammapp$,
with each point in the phase-space representing a possible
deformation condition within a small material region.
The independent variable is the vertical
position $x_{2}$ rather than time,
as would be the case in the usual
phase-space representations of oscillatory phenomena.
In a phase-plane of \mbox{$\gamma$ -- $\Gammap$}
(as in Fig.~\ref{fig:phase_spatial}),
uniform shear is simply a single point on the $\gamma$ axis; 
a strain discontinuity is represented by two separated points;
and a smooth shear band is represented by a loop or, perhaps, 
by a broken loop.
In the absence of horizontal body forces,
these phase trajectories represent the locus of combinations
$\gamma$, $\Gammap$, and $\Gammapp$ that would produce the same 
shearing stress $\sigma_{21}$ on horizontal planes.
We are
particularly interested in the possible presence of loop trajectories
within a family (phase portrait) 
of constant-stress trajectories,
as such loops indicate that the material could
support a shear band.
Our experiments allow the mapping of large portions of the phase portrait
for the assembly of 10,020 particles,
as will be seen in later sections.
\subsection{Testing series}
\begin{figure*}
\centering
\includegraphics{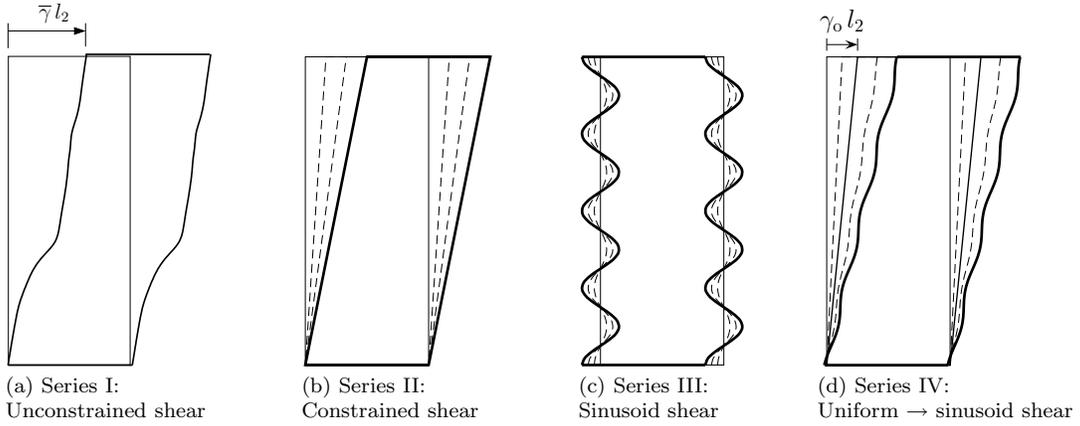}
\caption{Horizontal displacement contours in the four series of tests.
In Series II--IV, the deformations were coerced (constrained) by the
application of body forces.}
\label{fig:series_all}
\end{figure*}

Four series of tests were 
performed on the particle assembly, either by 
permitting unconstrained deformation or by coercing the
assembly through prescribed sequences of predetermined shapes
(Fig.~\ref{fig:series_all}).
These tests can be separated into two types
having different purposes:
\begin{itemize}
\item
Series~I tests were conducted under \emph{unconstrained} conditions 
without body forces, 
so that localization patterns were free to spontaneously develop
(Fig.~\ref{fig:series_all}a). 
These patterns were observed and cataloged.
\item
Series~II, III, and~IV tests were conducted under coerced, 
\emph{constrained} conditions that disallowed 
shearing localization
(Figs.~\ref{fig:series_all}b--d).
Body forces were applied to attain predetermined sequences of shapes,
with the intent of observing and measuring the
effects of strain gradients on the shearing response.
\end{itemize}
\subsection{Series~I tests: Unconstrained shear} \label{sec:series_I_methods}
In Series~I tests,
the shearing was produced by slowly moving
the upper boundary horizontally
at a rate $\dot{\gamma}l_{2}$ relative to the lower
boundary (Fig.~\ref{fig:series_all}a).
Unlike the other series of tests, no predetermined shape was imposed upon the
assembly, so that shear bands and other localization patterns
could freely develop within the assembly.
Furthermore, no body forces were
applied to the particles, so that the horizontal shearing stress
$\overline{\sigma}_{21}$ was uniform throughout the height of the
assembly.
\subsection{Series~II tests: Constrained uniform shear}
\label{sec:series_II_method}
This series of tests coerced a 
\emph{uniform shearing profile}
along the entire height of the assembly,~$l_{2}$,
with the sinusoid parameter $B$ set to zero
(Eq.~\ref{eq:U1} and Fig.~\ref{fig:series_all}b).
Unlike the unconstrained Series~I tests, 
a uniformly deformed shape was attained
by applying body forces, which disallowed
the formation of shear bands within the assembly and minimized the
strain gradients~$\Gammap$ and~$\Gammapp$.
\subsection{Series~III tests:  Monotonic sinusoid shear}%
\label{sec:series_III_method}
In the Series~III tests, the deformation profile was entirely sinusoid,
with the shape parameter $A$ set to zero 
(Eq.~\ref{eq:U1} and Fig.~\ref{fig:series_all}c).
The Series~III tests intentionally produced non-uniform
deformations, with non-zero shearing gradients $\Gammap$ and~$\Gammapp$
along the height of the assembly.
This series included 50 tests having various
combinations of the shape parameters~$n$ and~$\phi$,
so that the separate effects of $\Gammap$ and~$\Gammapp$
could be distinguished (Eq.~\ref{eq:U1}).
The integer~$n$ ranged from 1 to 8,
with larger values producing deformations having greater
strain gradients.
The largest value of~$n=8$
produced~16 half-sinusoid deformation zones, which
might be likened to half-sinusoid shear bands of width
$l_{2}/16$, or about $8.4D_{50}$---somewhat 
thinner than typical shear band widths.
Tests conducted with the same~$n$ but with different phase angles~$\phi$
simply shifted the deformation pattern so that 64 to 128 samples of the
stress-strain behavior could be captured and averaged
along the height of the assembly.
\par
\begin{figure}
\centering
\includegraphics{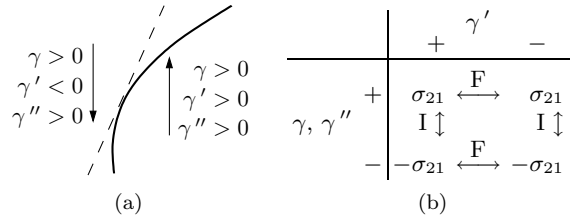}
\caption{Material frame indifference
and isotropy for a one-dimensional continuum.
The relations ``I'' and ``F'' are consequences
of isotropy and objectivity, respectively.}
\label{fig:frame_isotropy}
\end{figure}

In the Series~III tests, the shear strain and its
gradients were advanced in a monotonic and proportional manner, 
so that the ratios $\Gammap/\gamma$ and $\Gammapp/\gamma$
remained constant at any level $x_{2}$ within the assembly.
By taking advantage of material frame indifference (objectivity)
and isotropy, the behaviors at several levels $x_{2}$ can be
folded into a single quadrant of the 
$\gamma$ -- $\Gammap$ -- $\Gammapp$ phase-space.
We note that two observers, each rotated by
180$^{\circ}$ with respect to the other,
would report the same shearing deformation
$\gamma$, the same second strain gradient $\Gammapp$, 
and the same shearing stress $\sigma_{21}$
for a given deformation pattern (Fig.~\ref{fig:frame_isotropy}a).
They would, however, report opposite values of the first gradient 
$\Gammap$.
A one-dimensional constitutive form for the shearing stress must,
therefore, satisfy the condition 
\begin{equation}
\sigma_{21}(\gamma,\Gammap,\Gammapp) = \sigma_{21}(\gamma,-\Gammap,\Gammapp)\;,
\end{equation}
which implies an intrinsic nonlinearity in $\Gammap$.
When present in a constitutive form, the gradient $\Gammap$
should appear, therefore, as a norm, and, henceforth, the gradient
$\Gammap$ will often be written as the magnitude $|\Gammap|$.
If the material is isotropic, the shearing stress must also change
signs with a change in the signs of both $\gamma$ and $\Gammapp$:
\begin{equation}
\sigma_{21}(\gamma,\Gammap,\Gammapp) = -\sigma_{21}(-\gamma,\Gammap,-\Gammapp)
\;.
\end{equation}
These relationships are summarized in 
Fig.~\ref{fig:frame_isotropy}b.
\subsection{Series~IV tests:  Constrained incremental sinusoid shear}%
\label{sec:series_IV_method}
A fourth series of tests measured the
incremental response to loading and unloading in the
presence of strain gradients.
These tests allow us to study spontaneous localization
after a sustained period of uniform deformation.
The assembly was uniformly deformed
to predetermined 
shear strains $\gamma_{\mathrm{o}}$ (as in the Series~II tests),
and then it was deformed in small increments of non-uniform,
constrained sinusoid shearing (Fig.~\ref{fig:series_all}d).
By applying particular combinations of increments
$dA$ and $dB$ in Eqs.~(\ref{eq:U1})--(\ref{eq:gammapp}),
one of either $d\gamma$, $|d\Gammap|$, or $d\Gammapp$
could be advanced, while the other two would remain
stationary (Table~\ref{table:dA_dB}).
\begin{table}
\caption{Increments of $d\gamma$, $d\Gammap$, and $d\Gammapp$
applied in Series~IV tests after an initial
phase of uniform deformation.}
\label{table:dA_dB}
\centering\small
\renewcommand{\arraystretch}{1.25}
\begin{tabular}{cccc}
\hline
Condition &
$d\gamma$ & $|d\Gammap|$ & $d\Gammapp$ \\
\hline
1 & $+$ or $-$ &  0  &  0  \\
2 &  0  & $+$ &  0  \\
3 &  0  &  0  & $+$  or $-$ \\
4 & $+$ & $+$ & 0 \\
5 & $-$ & $+$ & 0 \\
\hline
\end{tabular}
\normalsize
\end{table}
These combinations were also employed
to reconcile possible incremental nonlinearities
associated with loading or unloading.
In addition to probing the separate incremental effects of
$d\gamma$, $|d\Gammap|$, and $d\Gammapp$, 
two other sets of tests were conducted
to investigate the possible coupling of the incremental effects
of strain $d\gamma$
and its first gradient $|d\Gammap|$
(Table~\ref{table:dA_dB}, Conditions~4 and~5).
\section{Results: localization and uniform shearing (Series~I and~II tests)}%
\label{sec:results}
\subsection{Localization patterns (Series~I tests)}%
\label{sec:ushear}
Localization was explored with a Series~I test, 
which produced shearing 
along the height of the assembly,
but under unconstrained conditions that allowed 
spontaneous localization
(Section~\ref{sec:series_I_methods} 
and Fig.~\ref{fig:series_all}a).
Three localization patterns were observed:
thin microbands at small strains, non-persistent
shear bands at moderate strains, and a solitary persistent shear band
just prior to and following the onset of strain softening.
\subsubsection{Microbands at small strains}\label{sec:microbands}
\begin{figure}
\centering
\includegraphics%
{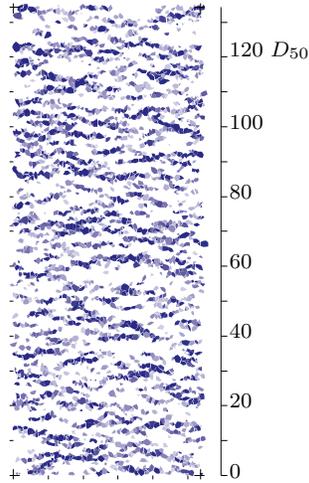}
\caption{Microband deformation patterns at small strains
in a Series~I test ($\gamma=0.01\%$).}
\label{fig:micro_bands}
\end{figure}

\emph{Microbands}, thin zones of intense shearing,
can appear at low strains, and even at the
very start of shear loading \cite{Kuhn:1999a}.
Figure~\ref{fig:micro_bands} shows the local shearing
deformations at the start of a Series~I unconstrained shear test.
The figure displays the horizontal shearing rates within
the smallest possible regions for which a deformation measurement
would be meaningful---the small polygonal void cells
within circuits of adjacent particles
(see \citeNP{Kuhn:1999a}).
Figure~\ref{fig:micro_bands} reveals a network of microbands.
Unlike the much thicker shear bands that appear 
at larger strains, microbands can be as short as $8D_{50}$,
with thicknesses
of $1$\textonehalf--$2$\textonehalf$D_{50}$.
The patterning is quite irregular, but the
center-to-center spacing between microbands is 
$3$\textonehalf--$8D_{50}$.
In Section~\ref{sec:gradients_microbands}, we relate this periodicity
to the measured dependence of shearing stress on the gradients of strain.
\subsubsection{Non-persistent and persistent shear bands at larger strains}
\label{sec:uncon}
\begin{figure*}
\centering\small
\includegraphics{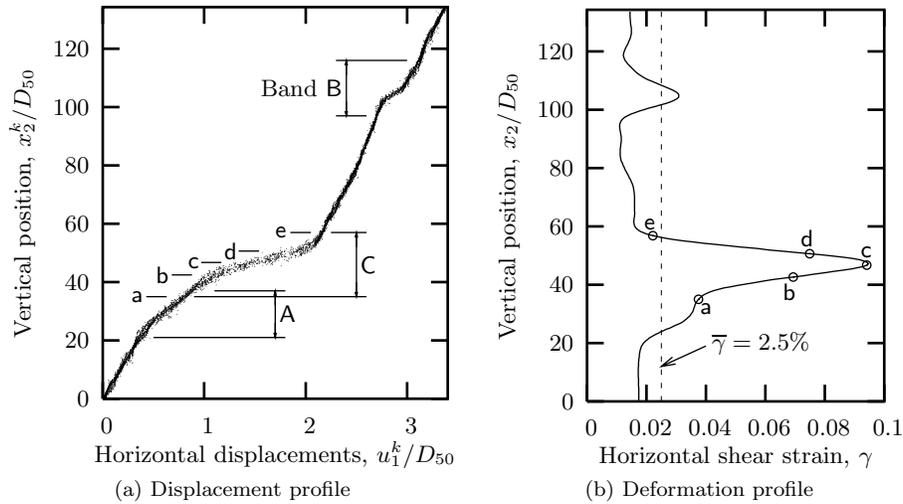}
\caption{Results of a Series~I test of unconstrained shear
for $\overline{\gamma}=0$ to $2.5\%$.
Three shear bands (labeled~$\mathsf{A}$, $\mathsf{B}$,~$\mathsf{C}$) 
appeared at various stages of deformation 
(see also Fig.~\ref{fig:compare_con_uncon}).  
Only band~$\mathsf{C}$
was eventually persistent.}
\label{fig:displacements_250}
\end{figure*}

Although the horizontal shearing stress was uniform along 
the assembly's height~$l_{2}$,
the shearing deformation was hardly uniform, as is apparent 
in Fig.~\ref{fig:displacements_250}, which
shows the horizontal displacements
$u_{1}^{k}$ of 10,020 particles at their
heights $x_{2}^{k}$ within the assembly.
In the figure, the shearing strain $\overline{\gamma}$ is 2.5\%,
and the particle displacements are measured relative to their initial positions
at $\gamma=0$.
Both the displacements~$u_{1}^{k}$ and 
positions~$x_{2}^{k}$ are expressed in a dimensionless
form by dividing by the median particle diameter~$D_{50}$.
\par
The residual evidence of three deformation bands is apparent in 
Fig.~\ref{fig:displacements_250}a.
Two of the bands, labeled~$\mathsf{A}$ and~$\mathsf{B}$,
were not persistent
and were no longer active at the 2.5\% strain; whereas, band~$\mathsf{C}$
had just become the sole and persistent shear band.
The non-persistent
bands~$\mathsf{A}$ and~$\mathsf{B}$ were present in varying intensities
at shear strains $\overline{\gamma}$ of 0.5--2.1\%
respectively, and each was, at times, the dominant deformation region.
At other times, each band was entirely inactive.
Such non-persistent bands have also been observed in biaxial tests
on loose sands \shortcite{Finno:1997a}.
Band~$\mathsf{C}$ first appeared at a strain of 1.4\% but was, at later
times, altogether inactive.  
It became the dominant and persistent 
band at a strain of 1.85\%, shortly before the onset of softening.
During the subsequent shearing, the assembly's deformation
was localized entirely within this single zone.
\par
In these experiments,
we see the truly complex influence of heterogeneity on shear
band formation,
where a local weakness can precipitate a shear band
but later heal, while yet other shear bands are forming
elsewhere within the material.
This transient behavior is also addressed in Sections~\ref{sec:variation} 
and~\ref{sec:model_of_bands}.
\par
The accumulated strains within the three bands are clearly 
present in the shear strain profile of Fig.~\ref{fig:displacements_250}b,
which gives shearing strains relative to the initial, undeformed assembly.
The persistent band~$\mathsf{C}$
has a thickness of about $20D_{50}$,
somewhat greater than that measured in other
experiments
(see~\shortciteNP{Vardoulakis:1998a} for a review).
\par
\begin{figure}
\centering\small
\includegraphics{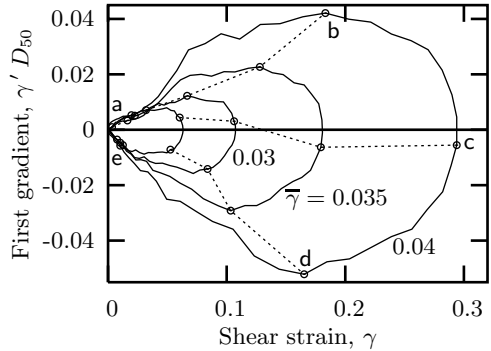}%
\caption{Phase-plane representation of 
shear band $\mathsf{C}$ during a Series~I unconstrained shear test.
Points $\mathsf{a}$--$\mathsf{e}$
are shown in Fig.~\ref{fig:displacements_250}.
Strains and strain gradients are
relative to those at $\gamma=0$.}
\label{fig:phase_plane_band}
\end{figure}

The deformations 
can also be represented in the phase-plane of $\gamma$ and $\Gammap$
(Fig.~\ref{fig:phase_plane_band}).
The loops in the figure are of band~\textsf{C}, which we have
highlighted by plotting the conditions at strain intervals of 0.5\%
relative to the reference strain of $\overline{\gamma}=2.0\%$.
The results 
show a progressive intensification of local shearing deformation within
the band,
as is indicated by the increasing size of the loop trajectories.
In this figure, the left ends of the loops cross the vertical
$\Gammap$-axis, although only slightly.
These negative strains indicate
a small unloading of the material outside of the shear band~$\mathsf{C}$.
In Section~\ref{sec:shear_band_gradients} we again 
consider a phase-space representation
of band~$\mathsf{C}$, 
by referencing a simple one-dimensional gradient-dependent model.
\subsection{Comparison of constrained and unconstrained shearing
(Series~I and~II)}%
\begin{figure}
\centering\small
\includegraphics{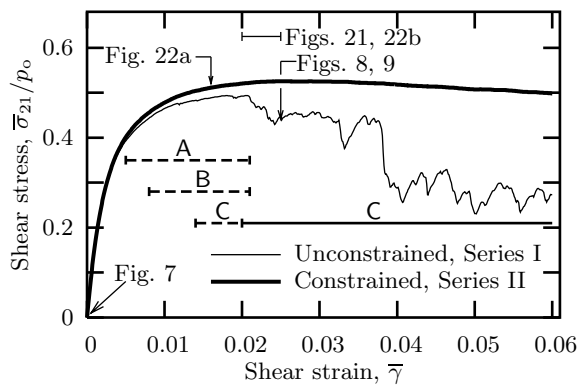}%
\caption{Comparison of constrained and unconstrained shearing tests
(Series~I and~II, Figs.~\ref{fig:series_all}a--b).
The three shear bands $\mathsf{A}$, $\mathsf{B}$,
and~$\mathsf{C}$ are shown in Fig.~\ref{fig:displacements_250}.}
\label{fig:compare_con_uncon}
\end{figure}

\label{sec:compare_con_uncon}
Shearing stresses for both constrained and unconstrained conditions
are compared in Fig.~\ref{fig:compare_con_uncon}.
During a constrained Series~II test, uniform shearing 
was coerced along the height of the assembly, 
using body forces to
preclude the appearance of shear bands
and microbands
(Section~\ref{sec:series_II_method} and Fig.~\ref{fig:series_all}b).
A constrained test can be thought to measure
the average material response, since a gross
deformation field was imposed upon the material, disallowing any spontaneous
localization within weaker regions.
The stresses at low strains are about equal for unconstrained and
constrained conditions, but the
stresses begin to differ at a strain
of about 0.4\%, which roughly coincides with the first
observation of a non-persistent shear band (band~$\mathsf{A}$,
Fig.~\ref{fig:displacements_250}).
The strengths in Fig.~\ref{fig:compare_con_uncon} abruptly diverge 
at a strain of 2.1\%, which roughly coincides with
the final emergence of band~$\mathsf{C}$ as the 
solitary persistent shear band.
The rather erratic strength at strains
greater than 3.0\% is likely due to discrete deformation events that involve,
perhaps, a few particles becoming either dislodged
or jammed among neighboring particles.
Shear band~$\mathsf{C}$ encompasses fewer than 1500 particles,
and such discrete deformation events would likely
have a dominant influence among particles that
are moving from one meta-stable configuration to another.
\section{Results: strain gradient effects for constrained 
proportional sinusoid shearing (Series~III tests)}\label{sec:results_III}
We now consider the results of Series~III tests in
which the assembly was progressively coerced into a sinusoid shape
(Section~\ref{sec:series_III_method} and 
Fig.~\ref{fig:series_all}c).
These tests will directly resolve whether the material
is simple or non-simple by scrutinizing the 
influence of the two gradients of shear strain, 
$\Gammap$ and $\Gammapp$,
under conditions in which $\gamma$, $\Gammap$, and $\Gammapp$
were proportionally and monotonically advanced.
The incremental response is probed with Series~IV tests
(Section~\ref{sec:results_IV}).
\subsection{Effect of second gradient $\Gammapp$ at large strains}
\label{sec:gammapp_effect}
The effect of the second gradient~$\Gammapp$ can be isolated
by comparing tests 
with different values of the shape
parameter~$n$ (see Eqs.~\ref{eq:U1}--\ref{eq:gammapp}).
Larger~$n$ values displace the assembly
into sinusoid profiles having more tightly spaced cycles with
larger gradients $\Gammapp$.
Although both~$\gamma$ and~$\Gammapp$ increase as the assembly is
progressively deformed,
the dimensionless ratio~$\Gammapp D_{50}^{2}/\gamma$ 
remains constant throughout the assembly and depends only upon the 
value of~$n$:
\begin{equation}\label{eq:gammappratio}
\frac{\Gammapp D_{50}^{2}}{\gamma} \;=\; 
-\left( \frac{2\pi n}{l_{2}/D_{50}}  \right)^{2}
\end{equation}
as in Eqs.~\ref{eq:gamma} and~\ref{eq:gammapp}
with $A=0$.
With values of~$n$ ranging from~1 to~8, we can explore the material's response
to a range of negative
ratios $\Gammapp D_{50}^{2}/\gamma$,  from nearly zero to 
about\ $-0.140$.
This range of ratios
fully encompasses the \emph{negative} gradients~$\Gammapp$ that would
be expected near the center of a shear band. 
We should note, however, that in these Series~III tests, the
ratio~(\ref{eq:gammappratio}) 
is always negative, so that the tests could not explore the effects of
positive ratios $\Gammapp/\gamma$,
which will occur along the outer portions of
a shear band (Section~\ref{sec:gradients_in_band}).
\par
\begin{figure}
\centering\small
\includegraphics{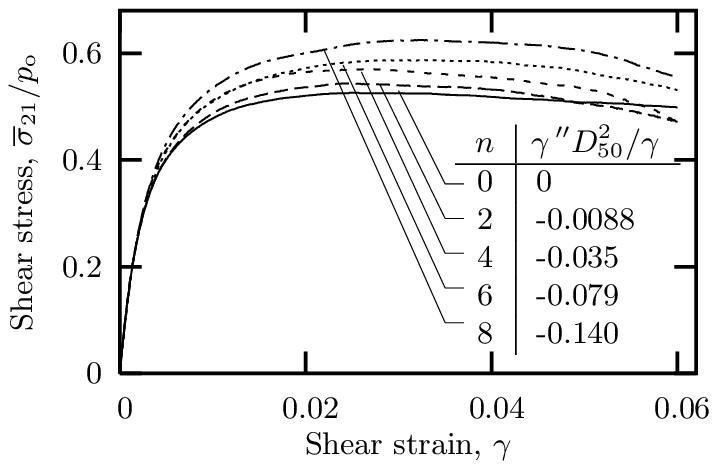}%
\caption{The measured effect of the second gradient of shearing
strain, $\Gammapp$, on the stress response of the material.}
\label{fig:Gammapp}
\end{figure}

The measured effect of the second gradient~$\Gammapp$ is shown in
Fig.~\ref{fig:Gammapp}.
The shearing stress $\overline{\sigma}_{21} /p_{\mathrm{o}}$
and the shearing strain~$\gamma$ are plotted for various
ratios of $\Gammapp D_{50}^{2}/\gamma$ (Eq.~\ref{eq:gammappratio}).
Each line is an average of as many as 128 samples that were subjected to the
same deformation sequence.
The single solid line for ratio $\Gammapp D_{50}^{2}/\gamma=0$ 
is borrowed from
the Series~II tests in which the assembly was uniformly sheared
under constrained conditions 
(Sections~\ref{sec:series_II_method} and~\ref{sec:compare_con_uncon}).
For a classical, simple continuum, the stress-strain response is
independent of $\Gammapp$.
The data shows, however, that
a negative second gradient~$\Gammapp$ has a moderate
hardening effect on granular materials, increasing the
local yield strength.
Figure~\ref{fig:Gammapp} shows the averaged results for all locations $x_{2}$
at which the first gradient~$\Gammap$ was equal to zero, so
that the effect of the second gradient~$\Gammapp$ is entirely isolated.
This particular condition of a negative second gradient~$\Gammapp$
and a first gradient~$\Gammap=0$ would be expected to occur
at the mid-thickness of a developing shear band (at point~``$\mathsf{c}$''
in the shear band~$\mathsf{C}$ of 
Figs.~\ref{fig:displacements_250} and~\ref{fig:phase_plane_band}).
Increasingly negative values of the ratio
$\Gammapp D_{50}^{2}/\gamma$ produce a larger hardening
effect, although even with low ratios,
the hardening effect is still active.
The smaller ratios correspond to half-sine thicknesses of 30 or more particle
diameters, indicating that the hardening effect of the second gradient
$\Gammapp$ may be active across quite large distances.
The lines in Fig.~\ref{fig:Gammapp} appear to merge 
at small strains, but subtle differences are present in the initial
slopes, 
and these differences
can be attributed to the second gradient $\Gammapp$,
a phenomenon that is
viewed more closely in Sections~\ref{sec:gammapp_small} 
and~\ref{sec:gradients_microbands}.
\subsection{Effect of first gradient~$\Gammap$ at large strains}
\label{sec:gammap_effect}
The Series~III tests also resolved the 
separate effect of the first gradient of shear strain.
A strong softening effect of the gradient~$\Gammap$ is revealed by
comparing the
material's response at different locations $x_{2}$ within the assembly.
Equations~(\ref{eq:gamma}) and~(\ref{eq:gammap}) 
can be arranged into an expression
for the first strain gradient,
\begin{equation} \label{eq:gammap_n_gamma}
\frac{\Gammap D_{50}}{\gamma} \;=\; \frac{2\pi n}{l_{2}/D_{50}}\cot
\left( \frac{2\pi n}{l_{2}} x_{2} - \phi \right) \, ,
\end{equation}
noting that shape parameter $A=0$ in the Series~III tests.
For a test in which the parameters~$n$ and~$\phi$ have been prescribed,
the dimensionless ratio~$\Gammap D_{50} / \gamma$
depends only on the position~$x_{2}$.
Because of spatial symmetries in the sinusoid displacement profile,
the averaged results can be presented in the form of the positive
ratio $\left\vert \Gammap D_{50} / \gamma \right\vert$
(Section~\ref{sec:series_III_method}).
\par
\begin{figure}
\centering\footnotesize
\includegraphics{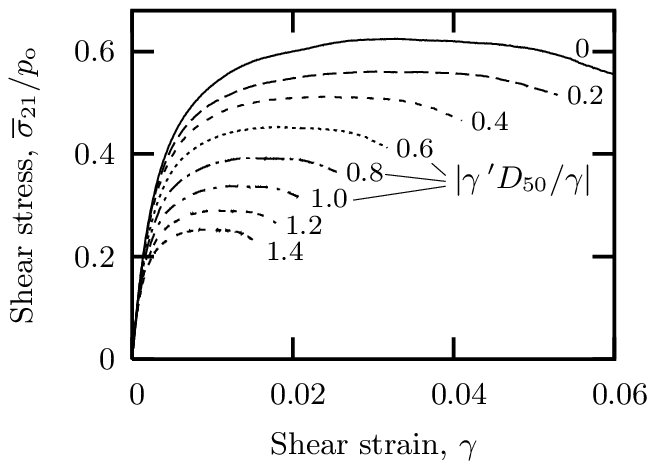}%
\caption{The measured effect of the first gradient of shear
strain, $\Gammap$, on the stress response of the granular material.}
\label{fig:Gammap_1}
\end{figure}

Figure~\ref{fig:Gammap_1} shows the averaged stress response at various ratios
$\left\vert \Gammap D_{50} / \gamma \right\vert$ for
tests in which $n=8$.
Each line shows the response to a different ratio
$\left\vert \Gammap D_{50} / \gamma \right\vert$ and is the averaged
behavior of as many as 128 material samples.
The figure documents a rather substantial softening effect of the first
gradient $\Gammap$ under monotonic loading conditions
in which $\gamma$, $\Gammap$, and $\Gammapp$
progress in a proportional manner.
\par
In the previous section, a negative second gradient $\Gammapp$
was shown to have a moderate hardening influence on
granular materials.
This influence was apparent in Fig.~\ref{fig:Gammapp} for
conditions with $\Gammap=0$.
At even moderate ratios $|\Gammap D_{50} / \gamma|$ of 0.4 or greater,
the hardening effect of a negative $\Gammapp$ is, however,
entirely counteracted by the softening influence of
the first gradient $\Gammap$.
\par
\begin{figure}
\centering\small
\includegraphics{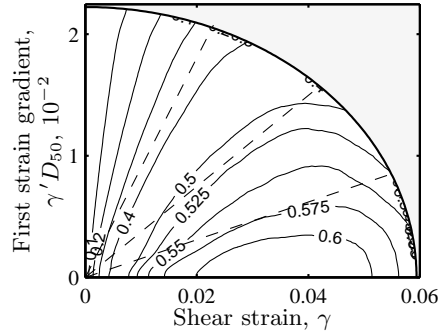}%
\caption{Contours of the
shearing stress $\overline{\sigma}_{21} /p_{\mathrm{o}}$.
The results are from tests in which $\gamma$ and $\Gammap$ were
increased in a proportional manner, as indicated by the dashed radial lines
(Series~III, $n=8$, $\Gammapp D_{50} / \gamma=-0.140$).}
\label{fig:Gammap_3}
\end{figure}

The contour plot of Fig.~\ref{fig:Gammap_3} shows the
combined effects of both the shear strain~$\gamma$ and its dimensionless
gradient $|\Gammap D_{50}|$ upon the shear stress
$\overline{\sigma}_{21}/p_{\mathrm{o}}$.
Because of spatial symmetries in the Series~III tests,
the full phase-plane can be folded and averaged into
a single quadrant (Section~\ref{sec:series_III_method}).
Figure~\ref{fig:Gammap_3} displays looped contours 
for stress ratios $\overline{\sigma}_{21} /p_{\mathrm{o}}$
in the range of about 0.55--0.62,
which resemble the looped phase trajectories of the shear bands that were 
depicted in Figs.~\ref{fig:phase_spatial}c--d.
Although the looped contours in Fig.~\ref{fig:Gammap_3}
might support the existence of a shear band,
the contours, by themselves, give no indication of how
a band could spontaneously form and then intensify.
If a region is, at first, deforming uniformly,
as would be represented by a single point on
the $\gamma$ axis (Fig.~\ref{fig:phase_spatial}a),
the phase trajectory
must somehow open into a looped trajectory 
at the start of a band-like localization.
(Figs.~\ref{fig:phase_spatial}b, c, or d).
Such incremental transformations are experimentally investigated 
with Series~IV tests 
in Sections~\ref{sec:dgamma_large} and~\ref{sec:model_of_bands}.
\fontsize{11}{13.8}\selectfont
\subsection{Constitutive implications}\label{sec:constitutive}
Except in micro-polar formulations, 
the influence of the first strain gradient is usually
excluded in gradient-dependent constitutive proposals.
Two rationales have been advanced for negating the 
influence of the first gradient of strain.
One argument arises in efforts to derive the constitutive
behavior of a granular material by 
averaging of the influence of surrounding
particles upon a single representative particle
\cite{Koenders:1990a,Muhlhaus:1994a,Chang:1995b}.
If the derivation is to be independent of the delegate particle,
then the statistical distribution of its neighborhood
must exhibit a central symmetry: 
the particle's neighborhood in a direction $\mathbf{n}$ 
must be equivalent to that in the opposite direction
$-\mathbf{n}$.
The combined assumptions of central symmetry and a linear contact behavior
cancel the possible effects of the first and all other odd-ordered
gradients of strain.
A second argument arises when gradient-dependent formulations
are derived as approximations of non-local, integral-type constitutive
forms \shortcite{Bazant:1984a,Vardoulakis:1991a,Vermeer:1994a}.
In this approach, the stress at a material point $\mathbf{x}$
is expressed as a functional of the strain in a finite neighborhood
$\mathcal{B}$ of the point, 
usually as an averaged strain $\overline{\boldsymbol{\epsilon}}$.
The surrounding deformations are averaged with a weighting
kernel $\boldsymbol{\Phi}$,
\begin{equation}
\overline{\boldsymbol{\epsilon}}(\mathbf{x}) = 
\underset{\mathcal{B}}{\int} 
\boldsymbol{\Phi}(\mathbf{x}-\mathbf{x}\boldsymbol{'}) 
\boldsymbol{\epsilon}(\mathbf{x}\boldsymbol{'})\,dV \;,
\end{equation}
and the integrand is then approximated with a Taylor polynomial
in the strain $\boldsymbol{\epsilon}(\mathbf{x}\boldsymbol{'})$ 
and its gradients.
The combined assumptions of a centrally symmetric kernel,
with $\boldsymbol{\Phi}(\mathbf{u}) = \boldsymbol{\Phi}(-\mathbf{u})$,
and a symmetric region $\mathcal{B}$
cancel the influence of odd-ordered gradients of strain.
\par
\begin{figure}
\centering\small
\includegraphics{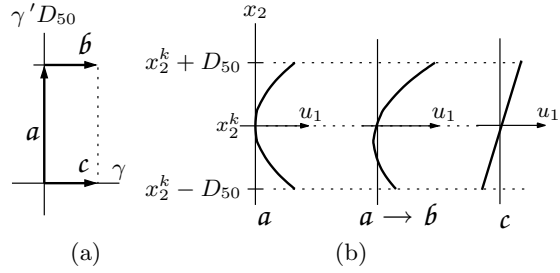}
\caption{Different deformation sequences and their relation to a material's
central symmetry.
}
\label{fig:central_symmetry}
\end{figure}

The strong measured influence of the first gradient $\Gammap$ 
in the current experiments
suggests that central symmetry is invalid when applied
to granular materials.
Two possible reasons for this invalidation can be surmised by considering
a single representative, $k^{\mathrm{th}}$, granule 
centered at a vertical position $x_{2}^{k}$ within
the assembly (Fig.~\ref{fig:central_symmetry}).
We can suppose a non-uniform deformation field in which the entire
assembly is deformed by horizontal movements
$u_1({x_2})$ that produce no shearing strain $\gamma$ at the level $x_{2}^{k}$
but produce a large
gradient of strain $\Gammap$.
These conditions near $x_{2}^{k}$ are represented with 
path~``\large$\mathpzc{a}$\normalsize''
in the phase-plane of Fig.~\ref{fig:central_symmetry}a.
We can also suppose that the reference granule is in contact 
with several neighboring
particles, including particles at $x_{2}^{k}+D_{50}$ and 
$x_{2}^{k}-D_{50}$.
If the relative movements of the three particles are large,
frictional slipping might occur at their contacts.
In granular materials, such inelastic response is accompanied
by significant changes in the material's fabric 
(e.g.,~\shortciteNP{Oda:1980a}).
Because of the opposite directions of shearing at the two locations,
fabric changes will be quite different at $x_{2}^{k}+D_{50}$ and 
$x_{2}^{k}-D_{50}$, and any initial central symmetry would be lost.
Suppose also that this first stage of deformation is followed
by a second stage that produces a small uniform shearing strain $\gamma$
at level $x_{2}$, as with path~``\large$\mathpzc{b}$\normalsize''
in Fig.~\ref{fig:central_symmetry}a.
Due to the nonlinear nature of the contact mechanism,
the previous stage~\large$\mathpzc{a}$\normalsize\ of non-uniform deformation
will alter the incremental response to the new shearing
deformation~\large$\mathpzc{b}$\normalsize.
The neighboring particle at position $x_{2}^{k}+D_{50}$
will likely continue to slip across the reference
particle; whereas, slipping would cease with the particle at 
$x_{2}^{k}-D_{50}$, as its
contact is unloaded by the \large$\mathpzc{b}$\normalsize\ displacements.
Because of this nonlinearity in the contact mechanism,
the central symmetry is further broken by the previous 
stage~\large$\mathpzc{a}$\normalsize\ of non-uniform deformation.
The response through 
stages~\large$\mathpzc{a}$\normalsize\ %
and~\large$\mathpzc{b}$\normalsize\ would 
likely differ, therefore, from that produced by uniform deformation alone
(for example, by the 
path~``\large$\mathpzc{c}$\normalsize'' in Fig.~\ref{fig:central_symmetry}).
The experimental results suggest that central symmetry is annulled
in a manner that produces substantial gradient-dependent behavior in granular
materials.
\fontsize{11}{13.8}\selectfont
\subsection{Dilatancy}\label{sec:dilatancy}
The material in this study dilated as it was sheared---a distinctive 
characteristic of granular materials.
During the constrained uniform shearing of Series~II
tests, the rate of volume increase was as large as 0.30 times the
shearing rate $\dot{\gamma}$.
With the sinusoid Series~III tests, we can investigate whether
the dilation rate depends on either of the two gradients~$\Gammap$
or~$\Gammapp$ under conditions in which~$\gamma$, $\Gammap$, and~$\Gammapp$
advance proportionally.
Although dilation does depend upon the shearing
strain $\gamma$,
the tests disclosed minimal dependence of the dilation rate upon 
either the first or second gradients of shear strain.
The dilation that
accompanies shearing is generally thought to contribute to the peak strength
of granular materials.
The current experiments show that the gradients
$\Gammap$ and $\Gammapp$ can affect the material's strength,
but not through any coupling between
the dilation rate and the gradients of strain.
The results also show that the large dilation within a shear band
is caused by the large strains, not by the gradients of strain.
\fontsize{11}{13.8}\selectfont
\subsection{Particle rotations}\label{sec:rotations}
In a Cosserat framework of generalized continua,
rotations and displacements are treated as independent fields, so
that the rotation of a material point can differ
from the mean rotation of its neighborhood
(i.e. the \emph{mean-field rotation}).
However,
simulations and physical experiments 
have shown that although individual particle rotations
can differ greatly from the mean rotation field,
the mean particle rotation is very close to 
the mean-field rotation (e.g., \shortciteNP{Calvetti:1997a}).
The current work provides an opportunity to test whether
the \emph{mean rotation field} differs from the \emph{mean-field rotation}
under conditions of non-uniform deformation.
\par
\begin{figure}
\centering\small
\includegraphics{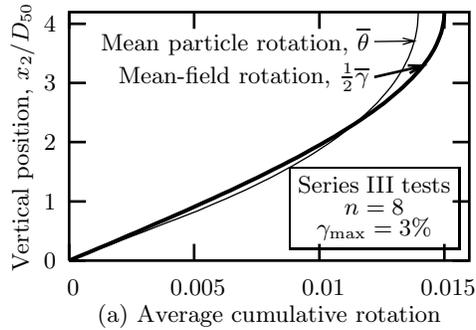}%
\caption{Comparison of the mean rotations $\overline{\theta}$
of individual particles and the local mean-field rotation
$\gamma/2$.}
\label{fig:rotations}
\end{figure}

The two fields are compared in Fig.~\ref{fig:rotations}, using data from
four Series~III tests with $n=8$, in which
the gradients of strain were quite large.
By running multiple tests, the mean rotation 
field $\overline{\theta}(x_{2})$ in Fig.~\ref{fig:rotations} was averaged 
and smoothed from
the rotations of about 40,000 sample particles.
The figure shows that the
mean rotation field and the mean-field rotation are nearly
equal. 
Any small differences likely result 
from the averaging of particle rotations that exhibit 
considerable variation:
the standard deviation of particle spins is remarkably about 12 
times the mean-field spin.
These large variations in particle rotation underscore the difficulty
of ascribing continuum behavior to discrete systems
of particles.
\fontsize{11}{13.8}\selectfont
\subsection{Spatial variations in the shearing response}\label{sec:variation}
\begin{figure}
\centering\small
\includegraphics{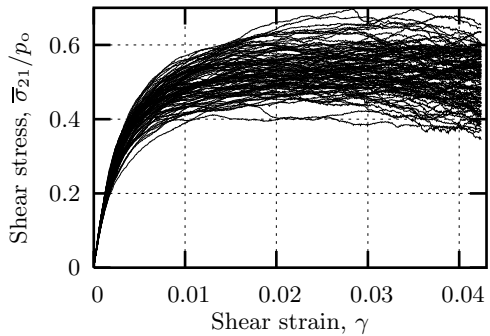}%
\caption{The spatial variation in 
strengths among 96 samples in a set of Series~III tests.}
\label{fig:scatter_shear}
\end{figure}

Until now, we have only considered the averaged material response.
Localized deformation likely initiates in locally weak regions,
as will be demonstrated in Section~\ref{sec:model_of_bands}, and
we currently consider the extent of these spatial variations in strength.
The responses of~96 samples to 
the same $\gamma$ -- $\Gammap$ -- $\Gammapp$ 
loading path
are shown in Fig.~\ref{fig:scatter_shear}.
For each sample, the shear strain and its gradients advanced
in a proportional manner, with the ratios 
$\left\vert\Gammap D_{50} / \gamma \right\vert$
and $\Gammapp D_{50}^{2} / \gamma$ remaining 0.20 and $-0.079$
respectively.
\par
Figure~\ref{fig:scatter_shear} reveals a rather large variation in the local
response among the 96 samples, each containing about~70 particles.  
In this respect, the local behavior
could be more properly referred to as a ``range of behaviors.''
The range became larger with increasing strain, but was significant even
at very low strains. 
\par
The many lines in Fig.~\ref{fig:scatter_shear} are highly intertwined,
so that the ordering of strengths among the 96 samples was constantly being
rearranged.
Zones that were weaker at one instant could later heal and strengthen.
This lack of temporal coherence can be quantified by comparing
the 96 strengths at two different strains, say 1\% and 4\%, for which
the coefficient of correlation was only 0.54.
\section{Results: strain gradient effects during incremental sinusoid shear 
(Series~IV tests)} \label{sec:results_IV}
The tests in the previous section were conducted with loading that
was monotonic and proportional.
In this section, we consider the \emph{incremental behavior} exhibited in 
Series~IV tests, in which a sustained stage of constrained uniform shearing
was followed by increments of constrained sinusoid shearing
(see Sections~\ref{sec:series_IV_method} and Fig.~\ref{fig:series_all}d).
Such experiments will aid in constructing a gradient-dependent flow
theory and in investigating the possible development of
localization bands after a sustained period of uniform shearing,
as in Section~\ref{sec:shear_band_gradients}.
We begin by addressing the incremental behavior at
the start of loading, and then
we consider the incremental response after an extended stage of uniform
deformation.
\fontsize{11}{13.8}\selectfont
\subsection{Incremental effects 
$|\Gammap|$ and $\Gammapp$ at small strains}
\label{sec:gammapp_small}
At the start of loading, the first strain
gradient $\Gammap$ had no discernible effect upon the material
stiffness.
This observation suggests that, on average, the local material
behavior exhibits the sort of central symmetry, 
discussed in Section~\ref{sec:constitutive},
although this symmetry subsequently disappears at larger strains.
The central symmetry at small strain is consistent with the conditions that
were present in the initial particle configuration:
the initial particle arrangement was isotropic, and the
particle contact behavior was initially elastic.
\par
\begin{figure}
\centering\small
\includegraphics{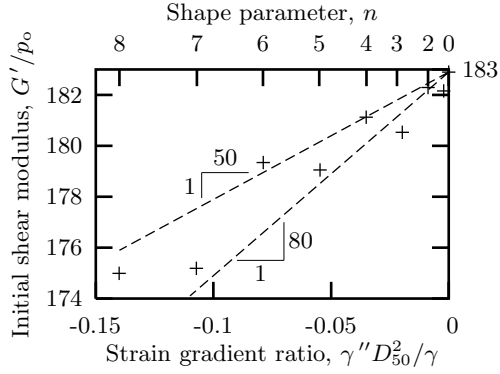}%
\caption{The effect of the second gradient of shear strain, $\Gammapp$,
on the initial, small-strain shear modulus $G'$.}
\label{fig:G_gammap_pp}
\end{figure}

At small strains,
a negative gradient $\Gammapp$ \emph{reduces} the material's
initial stiffness.
This behavior is opposite that observed
at large strains, where an increasingly negative second gradient $\Gammapp$
was found to have a cumulative 
hardening effect (Section~\ref{sec:gammap_effect}).
The small strain behavior is
detailed in Fig.~\ref{fig:G_gammap_pp}, which shows 
the initial shear modulus
and its dependence on the ratio $\Gammapp D_{50}^{2}/\gamma$.
In the figure, the apparent small-strain shear modulus
$G' \equiv \partial\overline{\sigma}_{21} / \partial{\gamma}$
has been normalized by dividing by the initial pressure
$p_{\mathrm{o}} \equiv -$\textonehalf$\overline{\sigma}_{ii}$.
Each of 
the nine data points corresponds to a single value of the shape
parameter $n$, ranging from~0 to~8 
(Eqs.~\ref{eq:U1}--\ref{eq:gammapp}),
and each data point represents the averaged response of as many as 128
material samples. 
The data was collected in locations $x_{2}$ at which 
the first strain gradient $\Gammap$ was zero
(as in Fig.~\ref{fig:Gammapp}).
The results in Fig.~\ref{fig:G_gammap_pp} are approximated as
\begin{equation}\label{eq:G_gammapp}
G' \approx G_{\mathrm{o}} + B_{2}(\Gammapp/\gamma) \;,
\end{equation}
where $B_{2}$ is between $50 D_{50}^{2} p_{\mathrm{o}}$ and 
$80 D_{50}^{2} p_{\mathrm{o}}$,
and the reference shear modulus $G_{\mathrm{o}}$ is about $183p_{\mathrm{o}}$.
\fontsize{11}{13.8}\selectfont
\subsection{Incremental effects of 
$\gamma$, $|\Gammap|$, and $\Gammapp$ at large strains}%
\label{sec:dgamma_large}
Series~IV tests were used to measure the separate incremental
effects of 
shear strain and its gradients at larger strains
(Fig.~\ref{fig:series_all}d). 
The increments $d\gamma$, $|d\Gammap|$, and $d\Gammapp$
were applied after an initial stage
of steady uniform shearing to seven different predetermined 
strains $\gamma_{\mathrm{o}}$, ranging from $0.5\%$ to $5\%$
(Section~\ref{sec:series_IV_method}).
The tests reveal that the incremental response
depends upon each of the increments $d\gamma$, $|d\Gammap|$, and $d\Gammapp$.
The measured response of the shearing stress $d\overline{\sigma}_{21}$
can be approximated in the following form:
\begin{equation}\label{eq:dgammapp_ep}
d\overline{\sigma}_{21} \approx
G^{\mathrm{ep}} d\gamma \:+\:
B_{1}^{\mathrm{ep}} \left| d\Gammap  \right| \:+\:
B_{2}^{\mathrm{ep}} d\Gammapp \;.
\end{equation}
Measured values of the elasto-plastic moduli
$G^{\mathrm{ep}}$, $B_{1}^{\mathrm{ep}}$, and $B_{2}^{\mathrm{ep}}$
are given in Table~\ref{table:dgammapp_ep},
noting that incremental non-linearities in the first and third
terms of~(\ref{eq:dgammapp_ep}) are expressed with different
loading and unloading stiffness for
$G^{\mathrm{ep}}$ and $B_{2}^{\mathrm{ep}}$.
\begin{table}
\caption{Experimentally measured incremental stiffnesses for 
Eq.~\ref{eq:dgammapp_ep}.}
\label{table:dgammapp_ep}
\centering\small
\renewcommand{\arraystretch}{1.00}
\begin{tabular}{p{0.0cm} l  l  c}
\hline
\multicolumn{2}{c}{Stiffness} & Conditions & Values$^{a}$ \\
\hline
& $G^{\mathrm{ep},+}$     & Loading, $d\gamma / \gamma_{\mathrm{o}} > 0$   & 
  Fig.~\ref{fig:G_load_unload} \\
& $G^{\mathrm{ep},-}$     & Unloading, $d\gamma / \gamma_{\mathrm{o}} < 0$ & 
  Fig.~\ref{fig:G_load_unload} \\
& $B_{1}$                 & $d\gamma = 0$ & $-14D_{50} p_{\mathrm{o}}$\\
&           & $|d\Gammap|D_{50}/d\gamma = -0.4$ & $-11D_{50} p_{\mathrm{o}}$\\
&           & $|d\Gammap|D_{50}/d\gamma =  0.4$ & $ -3D_{50} p_{\mathrm{o}}$\\
& $B_{2}^{\mathrm{ep},+}$ & $d\Gammapp / \gamma_{\mathrm{o}} >0$, $d\gamma = 0$           & $-100D^{2}_{50}p_{\mathrm{o}}$\\
& $B_{2}^{\mathrm{ep},-}$ & $d\Gammapp / \gamma_{\mathrm{o}} <0$, $d\gamma = 0$           & $-16D^{2}_{50}p_{\mathrm{o}}$\\
\hline
\multicolumn{4}{l}{$^{a}$\ also see Fig.~\ref{fig:coupling_2}
for values of $B_{1}^{\mathrm{ep}}$}\\
\hline
\end{tabular}
\normalsize
\end{table}
The effects of the shearing strain
and its first two gradients are shown as being uncoupled
in~(\ref{eq:dgammapp_ep}). 
We will soon see, however, that the influences of $d\gamma$ and $|d\Gammap|$
are likely interdependent.
\par
\begin{figure}
\centering\small
\includegraphics{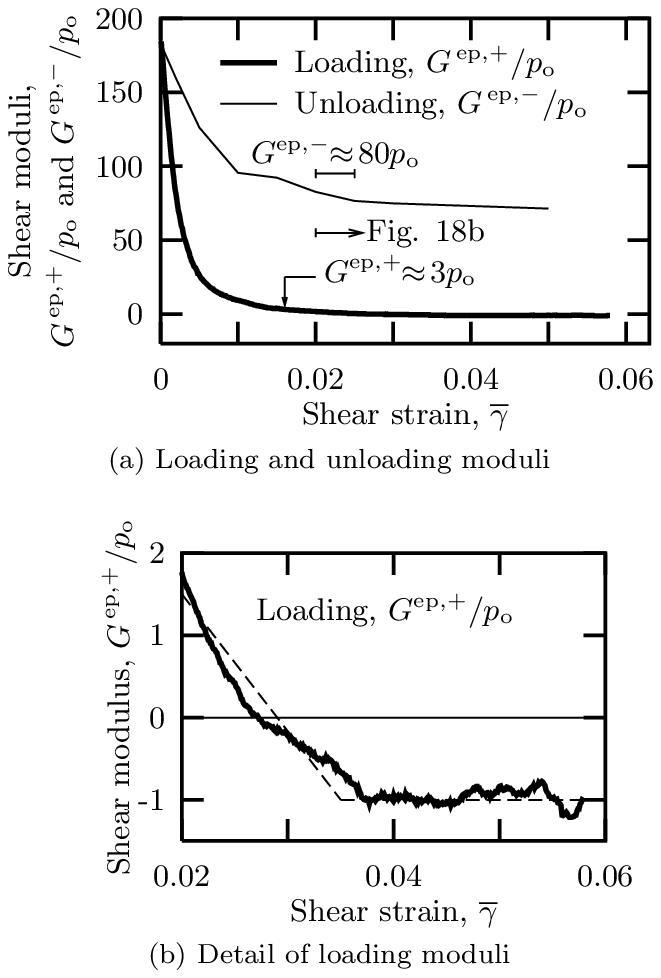}
\caption{The incremental shear moduli, for both loading and unloading, as
a function of the advancing shear strain.}
\label{fig:G_load_unload}
\end{figure}

During loading ($+$), the 
elasto-plastic shear stiffness $G^{\mathrm{ep},+}$ is simply
the slope of the 
stress-strain curve in Fig.~\ref{fig:compare_con_uncon},
and this slope becomes slightly negative at strains 
larger than 2.7\% (Fig.~\ref{fig:G_load_unload}).
The \emph{unloading stiffness} $G^{\mathrm{ep},-}$ was continually degraded
during the loading process.
\par
A negative second strain gradient $d\Gammapp$ has an incremental 
stiffening effect on the material, with an associated negative modulus
$B_{2}^{\mathrm{ep}}$.
Although this influence is consistent with the 
cumulative effect of 
$\Gammapp$ during proportional loading
(Section~\ref{sec:gammapp_effect}), it is opposite
the incremental effect at small strain, where $B_{2}$ is positive
(Sections~\ref{sec:gammapp_effect} and~\ref{sec:gammapp_small}).
The incremental stiffness $B_{2}^{\mathrm{ep}}$ is discontinuous:
the value of $B_{2}^{\mathrm{ep}}$ is more negative when
$d\Gammapp$ is in the same direction as the
cumulative shear strain $\gamma_{\mathrm{o}}$ 
($d\Gammapp / \gamma_{\mathrm{o}} > 0$)
than when the increment $d\Gammapp$ is in the opposite direction
($d\Gammapp/\gamma_{\mathrm{o}} < 0$).
The two stiffnesses are labeled $B_{2}^{\mathrm{ep},+}$
and $B_{2}^{\mathrm{ep},-}$.
At the center of a developing shear band, 
the ratio $d\Gammapp / \gamma_{\mathrm{o}}$
is negative (for example, between the points $\mathsf{b}$ and $\mathsf{d}$
in Fig.~\ref{fig:displacements_250});
whereas $d\Gammapp / \gamma_{\mathrm{o}}$ 
is positive within the outer portions of
a shear band  (within the ranges $\mathsf{a}$--$\mathsf{b}$ and 
$\mathsf{d}$--$\mathsf{e}$,
Fig.~\ref{fig:displacements_250}).
These results indicate that
the strengthening effect of $d\Gammapp$ will be smaller
at the center of a shear band than will its
weakening effect near the extremities of the shear band.
\par
\begin{figure*}
\centering\small
\includegraphics{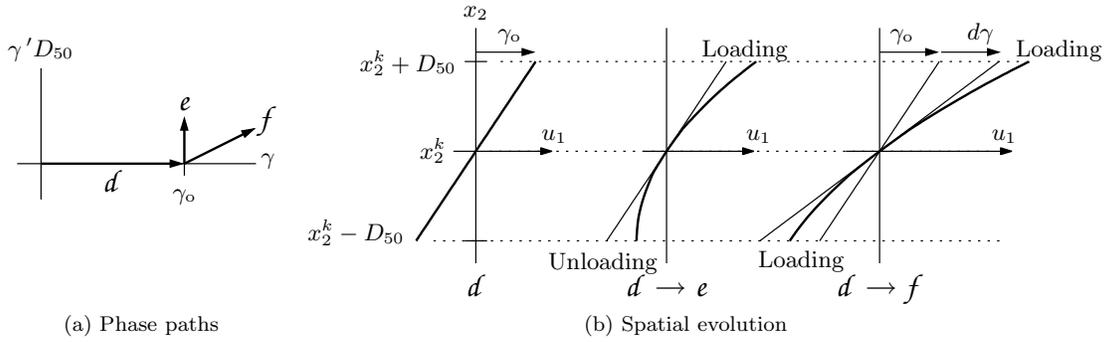}
\caption{%
A rationale for the possible coupling of $d\gamma$ and $|d\Gammap|$
in a constitutive form.
}
\label{fig:coupling_1}
\end{figure*}

The incremental stiffness $B_{1}^{\mathrm{ep}}$ 
that is associated with the first gradient
$|d\Gammap|$ is negative, which is consistent with the
cumulative weakening effect that was 
observed during the Series~III
tests (Section~\ref{sec:gammap_effect}).
The values of $B_{1}$ in Table~\ref{table:dgammapp_ep}
are averages of values that were measured after first uniformly
deforming the material
to seven different shear strains $\gamma_{\mathrm{o}}$ between 0.5\% and~5\%.
The influence of $|d\Gammap|$ at these
strains, altogether absent at strain $\gamma_{\mathrm{o}}=0$,
suggests that the material's initial central symmetry was broken
during the earliest stage of uniform loading.
This loss of symmetry should not be due to any stress-induced anisotropy
of fabric that would have been produced during the 
initial stage of uniform shearing, since a spatially uniform deformation
would have produced spatially uniform changes
in fabric.
The loss of central symmetry is probably due to 
non-linearities in the contact mechanism
that would develop during any inelastic deformation
(Section~\ref{sec:constitutive}).
Figure~\ref{fig:coupling_1} presents a rationale for this early loss
of central symmetry.
\fontsize{11}{13.8}\selectfont
The sequence \large$\mathpzc{d}\rightarrow\mathpzc{e}$\normalsize\ is
a uniform deformation 
``\large$\mathpzc{d}$\normalsize'' to a strain $\gamma_{\mathrm{o}}>0$,
followed by an increment 
``\large$\mathpzc{e}$\normalsize''
of purely non-uniform deformation.
\fontsize{11}{13.8}\selectfont
\fontsize{11}{13.8}\selectfont
In stage~\large$\mathpzc{e}$\normalsize, the increment $|d\Gammap|$ would
produce further loading, perhaps further sliding at
$x_{2}^{k}+D_{50}$, but would likely produce
\emph{unloading} in the vicinity of $x_{2}^{k}-D_{50}$.
This asymmetry would alter the resulting stress 
increment.
\fontsize{11}{13.8}\selectfont
\par
By extending this simple reasoning,
we would expect the stress response to depend upon a coupling of 
the increments $d\gamma$ and $|d\Gammap|$,
perhaps with a coefficient $B_{1}^{\mathrm{ep}}$ in
(\ref{eq:dgammapp_ep}) that depends upon 
the two increments $d\gamma$ and $|d\Gammap|$.
For example,
if the deformation was \emph{nearly} uniform, with a
substantial increment $d\gamma$ and a much smaller increment 
$|d\Gammap|$,
loading and sliding would likely continue at both of the locations
 $x_{2}^{k}+D_{50}$ and $x_{2}^{k}-D_{50}$
(see path~\large$\mathpzc{d}\rightarrow\mathpzc{f}$\normalsize, 
Fig.~\ref{fig:coupling_1}).
\fontsize{11}{13.8}\selectfont
Under these conditions,
any asymmetry produced by the sequence 
\large$\mathpzc{d}\rightarrow\mathpzc{f}$\normalsize\ would
probably be less than that produced by path
\large$\mathpzc{d}\rightarrow\mathpzc{e}$\normalsize,
with the latter producing a greater contribution from the
term $B_{1}^{\mathrm{ep}} |d\Gammap|$ in (\ref{eq:dgammapp_ep}).
\fontsize{11}{13.8}\selectfont
\par
\begin{figure}
\centering
\includegraphics{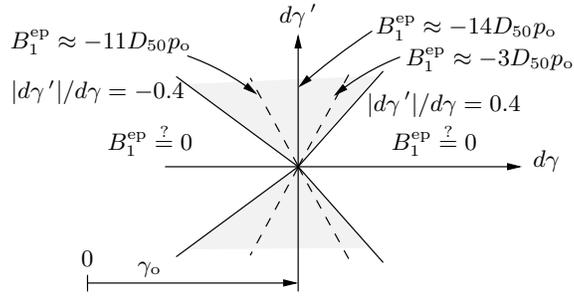}%
\caption{Experimental values of $B^{\mathrm{ep}}_{1}$ 
for different combinations of $d\gamma$ and $|d\Gammap|$
in Series~IV tests.}
\label{fig:coupling_2}
\end{figure}

This reasoning was examined with two sets of tests in which the 
increments $d\gamma$ and $|d\Gammap|$ were non-zero, and in which
$d\Gammapp$ was zero (Section~\ref{sec:dgamma_large} and Conditions~4 and~5
in Table~\ref{table:dA_dB}).
Stresses were measured at locations where the ratio 
$|d\Gammap D_{50}|/d\gamma$ was either $-0.4$ or $0.4$.
For both conditions,
the negative value of $B_{1}^{\mathrm{ep}}$ 
was smaller than its value with $d\gamma = 0$,
a trend that is consistent with the reasoning in the preceding paragraph.
The results are summarized in Fig.~\ref{fig:coupling_2},
which includes a certain speculation:
that the incremental effect of the first strain gradient
$|d\Gammap|$ is entirely negated at small ratios of $|d\Gammap|/d\gamma$.
\fontsize{11}{13.8}\selectfont
\section{Relationship of strain gradients and localization patterns}%
\label{sec:gradients_localization}
We now explore relationships between localization patterns
(Section~\ref{sec:results}) and the 
measured gradient-dependent behavior
(Sections~\ref{sec:results_III} and~\ref{sec:results_IV}).
To investigate these relationships, we construct a simple
one-dimensional gradient-dependent model that is consistent with
results of the constrained sinusoid tests.
We then compare the behavior of the model to the localization
patterns that were observed in the unconstrained shear tests.
We consider both microband and shear band localizations.
\fontsize{11}{13.8}\selectfont
\subsection{Microbands and strain gradients}\label{sec:gradients_microbands}
The experimental results in Section~\ref{sec:gammapp_small}
show that, at very small strains, shear stress
depends upon both the shearing strain $\gamma$ and its second
gradient $\Gammapp$.
We will see that this measured behavior is consistent
with the observed periodicity in microband patterning.
In the absence of body forces, a simple, one-dimensional model 
of shearing can be constructed from
the equilibrium and kinematic conditions,
\begin{equation}\label{eq:equilibrium}
d\overline{\sigma}_{21}/dx_{2} = 0 \;,\quad
du_{1}/dx_{2} = \gamma \;,
\end{equation}
where $\overline{\sigma}_{21}$ is the shear stress
at vertical position $x_{2}$,
and $u_{1}$ is the corresponding horizontal displacement.
The measured small-strain material behavior in~(\ref{eq:G_gammapp})
can be linearized as
\begin{equation}\label{eq:linear_small_strain}
\overline{\sigma}_{21} \approx 
G_{\mathrm{o}}\gamma + B_{2} \Gammapp \;.
\end{equation}
Substituting~(\ref{eq:linear_small_strain}) 
and~(\ref{eq:equilibrium}$_{2}$) into~(\ref{eq:equilibrium}$_{1}$)
leads to a harmonic fourth order ordinary differential equation
in the shearing displacement $u_{1}$,
\begin{equation}
B_{2} \,u_{1}^{\mathrm{iv}} 
\:+\: G_{\mathrm{o}} u_{1}^{\prime\prime} = 0 \;,
\end{equation}
where differentiation is with respect to the vertical position $x_{2}$.
Although an affine solution $u_{1} = a + bx_{2}$ may apply
to an ideal, homogeneous material undergoing uniform shearing,
local perturbations due to material inhomogeneity will likely produce
periodic solutions having the
spatial periodicity $\lambda = 2 \pi \sqrt{B_{2}/G_{\mathrm{o}}}$.
Experimental values of the stiffnesses $B_{2}$ and $G_{\mathrm{o}}$ 
were measured at the start of the Series~III sinusoid shear tests,
and these values
suggest a periodicity in the range of 3.3$D_{50}$ to 4.2$D_{50}$
(see Fig.~\ref{fig:G_gammap_pp}).
This range overlaps the observed range of periodicities of
3.5$D_{50}$ to~8$D_{50}$ for the microband patterning that
emerged at the very start of unconstrained shearing
(Section~\ref{sec:microbands} and Fig.~\ref{fig:micro_bands}).
These results suggest that a gradient-dependent continuum model
may account for certain features of micro-level localization.
\fontsize{11}{13.8}\selectfont
\subsection{Shear bands and strain gradient dependence}%
\label{sec:shear_band_gradients}
We now consider whether the observed features of shear bands
are consistent with the measured effects of 
strain gradients.
We will be primarily interested in the incremental effects of
the shear strain $\gamma$ and its first two derivatives,
$\Gammap$ and $\Gammapp$, as reported
in Section~\ref{sec:dgamma_large}.
We will relate these results to the shear band observations
of Section~\ref{sec:uncon}
by employing two separate approaches:
\begin{itemize}
\item
We first consider internal shear band features:  specifically,
we investigate the strains and strain gradients
within a shear band and determine whether these measurements are consistent
with the separately measured incremental effects of $d\gamma$, $|d\Gammap|$, 
and $d\Gammapp$ (Section~\ref{sec:gradients_in_band}).
\item
We also construct a simple one-dimensional model that is consistent with
the measured incremental effects of $d\gamma$, $|d\Gammap|$,
and $d\Gammapp$,
and then we test whether this model can reproduce the observed features
of a shear band (Section~\ref{sec:model_of_bands}).
\end{itemize}
\fontsize{11}{13.8}\selectfont
\subsubsection{Gradient-dependence within a shear band}%
\begin{figure}
\centering\small
\includegraphics{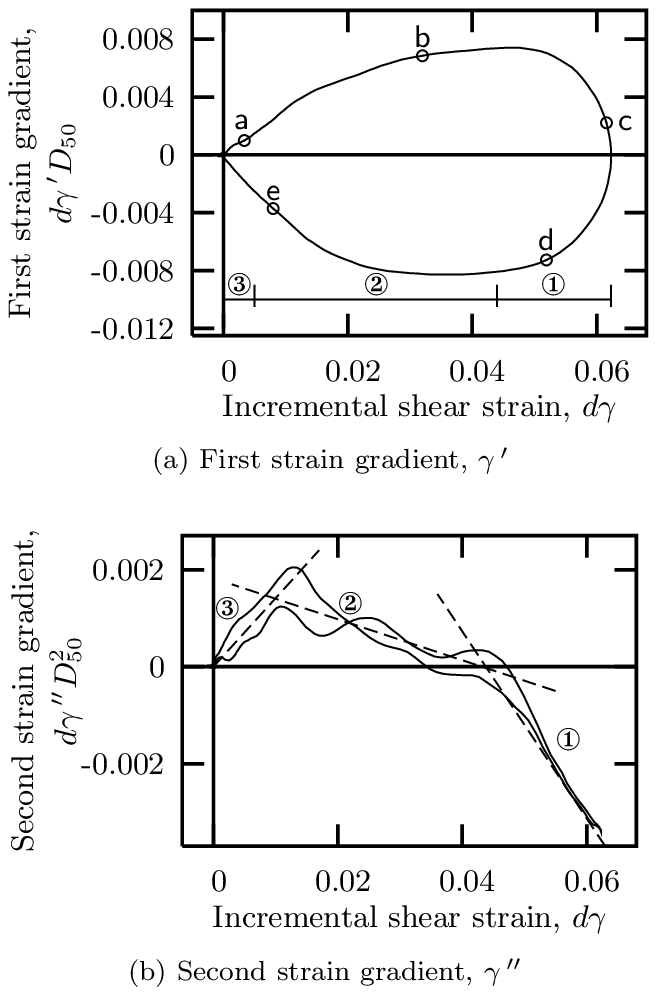}
\caption{The phase trajectory in 
\mbox{$\gamma$ -- $\Gammap$ -- $\Gammapp$}
space for the shear band~$\mathsf{C}$ in Fig.~\ref{fig:displacements_250},
during the strain interval of $2.0$--$2.5\%$.}
\label{fig:trajectory_200_250}
\end{figure}

\label{sec:gradients_in_band}
Figure~\ref{fig:trajectory_200_250} shows the phase trajectory 
for the persistent shear band~$\mathsf{C}$
(Section~\ref{sec:uncon} and Fig.~\ref{fig:displacements_250}).
Because the horizontal shear stress is uniform along the height of the
assembly, 
Fig.~\ref{fig:trajectory_200_250} gives 
combinations of the increments $d\gamma$, 
$|d\Gammap|$, and $d\Gammapp$ that produce the same increment
of shearing stress.
The deformations were measured 
between the average 
strains $\overline{\gamma}$ of 2\% and 2.5\%, after 
band~\textsf{C} had become persistent.
Although the average strain increased by $d\overline{\gamma}=0.5\%$,
the strains are substantially larger within the shear band,
where they have attained increments
as large as~6\%.
Within the 2.0--2.5\% strain interval, the material began to soften,
and the shearing stress decreased by $-0.053p_{\mathrm{o}}$
(see Fig.~\ref{fig:compare_con_uncon}).
\par
The phase plot of $d\gamma$ and $d\Gammapp$
appears to hold three regions with differing behaviors
(Fig.~\ref{fig:trajectory_200_250}b).
These regions are denoted with dashed trend lines and 
are labeled~\ding{172}, \ding{173}, and~\ding{174}.
Behavior~\ding{172} occurs near the center of the shear
band (near point~$\mathsf{c}$ in Fig.~\ref{fig:displacements_250}),
where shearing strains are large and the second strain gradient
$d\Gammapp$ is negative.
Behavior~\ding{173} occurs nearer the outer portions of the band, where
strains are smaller and the second gradient $d\Gammapp$
is positive.
The data shows a distinct change in slope
for behaviors~\ding{172} and~\ding{173}
(Fig.~\ref{fig:trajectory_200_250}b).
This change in slope is consistent with the incremental nonlinearity
associated with $d\Gammapp$, as was measured in the Series~IV
tests of Section~\ref{sec:dgamma_large}.
Indeed, if we ignore the effect of the first strain gradient
$|d\Gammap|$ in (\ref{eq:dgammapp_ep}),
the line~\ding{172} can be used to estimate the two moduli
$G^{\mathrm{ep},+}$ and $B_{2}^{\mathrm{ep},-}$ that are associated
with a positive, loading increment $d\gamma$ and a negative
increment $d\Gammapp$ ($d\gamma/\gamma_{\mathrm{o}}>0$, 
$d\Gammapp / \gamma_{\mathrm{o}} < 0$).
The stress increment $d\overline{\sigma}_{21}$ was 
$-0.053p_{\mathrm{o}}$, which leads to the following values
of $G^{\mathrm{ep},+}$ and $B_{2}^{\mathrm{ep},-}$:
\begin{equation}
G^{\mathrm{ep},+}     \approx -1.2 p_{\mathrm{o}} \;, \quad
B_{2}^{\mathrm{ep},-} \approx -6   D^{2}_{50} p_{\mathrm{o}} \;.
\end{equation}
The value of $G^{\mathrm{ep},+}$ within the
shear band is quite close to the loading
modulus of $-1.0p_{\mathrm{o}}$ that was measured 
at strains greater than $4\%$ 
(Fig.~\ref{fig:G_load_unload}b).
The computed sign of the modulus $B_{2}^{\mathrm{ep},-}$ is negative,
which is also consistent with that measured in the incremental
Series~IV tests of Section~\ref{sec:dgamma_large},
but the value $-6D^{2}_{50} p_{\mathrm{o}}$
is only about half that measured in the Series~IV tests
($B_{2}^{\mathrm{ep},-}$ in Table~\ref{table:dgammapp_ep}).
This difference may be the result of a possible coupling of the
increments $d\gamma$ and $d\Gammapp$ in their influence
on the stress increment $d\overline{\sigma}_{21}$.
The modulus $B_{2}^{\mathrm{ep},+}$ associated with
slope~\ding{173} is about $-30D^{2}_{50}p_{\mathrm{o}}$,
which also differs from the value measured
in Series~IV tests (Table~\ref{table:dgammapp_ep}).
\fontsize{11}{13.8}\selectfont
\subsubsection{One-dimensional model of shear bands}%
\label{sec:model_of_bands}
We now investigate shear bands by constructing
a simple one-dimensional gradient-dependent model that is consistent with the
experimental results of the Series~IV
sinusoid tests of Section~\ref{sec:dgamma_large}.
The incremental constitutive form (\ref{eq:dgammapp_ep}) that was constructed
from the experimental results is a non-homogeneous and
non-linear second order differential equation
in the incremental shearing strain $d\gamma$.
It is repeated here:
\begin{equation}
d\overline{\sigma}_{21} =
G^{\mathrm{ep}} d\gamma \:+\:
B_{1}^{\mathrm{ep}} \left| d\Gammap  \right| \:+\:
B_{2}^{\mathrm{ep}} d\Gammapp \;. \tag{\ref{eq:dgammapp_ep}}
\end{equation}
In the absence of body forces, the
horizontal shearing stress $d\overline{\sigma}_{21}$ must, of course,
be independent
of position $x_{2}$, as in~(\ref{eq:equilibrium}$_{1}$),
so $d\overline{\sigma}_{21}$ will be treated as an unknown parameter
that requires solving.
Equation~(\ref{eq:dgammapp_ep}) contains many complexities:
incremental non-linearities that are associated with the two
coefficients $G^{\mathrm{ep}}$ and $B_{2}^{\mathrm{ep}}$;
a derivative $d\Gammap$ that appears in an incrementally non-linear form;
and a coefficient $B_{1}^{\mathrm{ep}}$ that is a
non-linear function
of the ratio $|d\Gammap|/d\gamma$ (Section~\ref{sec:dgamma_large}). 
Our purpose is to test whether~(\ref{eq:dgammapp_ep}) 
can lead to incremental
solutions $du_{1}(x_{2})$ that approximate observed
shear band profiles.
To this end, we note that the horizontal shearing displacement $du_{1}$ 
is linked to~(\ref{eq:dgammapp_ep}) by the kinematic 
condition~(\ref{eq:equilibrium}$_{2}$).
We consider boundary conditions that would
apply to the upper half of a horizontal band centered at
$x_{2}=0$ within a domain $x_{2}\in [-L,L]$:
\begin{equation}\label{eq:boundary_all}
\begin{split}
du_{1}(L) = \overline{d\gamma}L = \overline{du}_{1}
\;,\quad
du_{1}(0) = 0\\
d\Gammap(0) = 0
\;,\quad
d\Gammap(L) = 0 \;.
\end{split}
\end{equation}
If the stress increment $d\overline{\sigma}_{21}(x_{2})$ is
constant, symmetries 
in~(\ref{eq:dgammapp_ep}), (\ref{eq:equilibrium}), (\ref{eq:boundary_all}) 
and the constitutive forms of $G^{\mathrm{ep}}$, 
$B_{1}^{\mathrm{ep}}$, and $B_{2}^{\mathrm{ep}}$ will impose the following
symmetry condition: $d\gamma(-x_{2}) = d\gamma(x_{2})$.
The height $L=67D_{50}$ will be half the height of the
particle assembly in Fig.~\ref{fig:assemblies},
and condition~(\ref{eq:boundary_all}$_{4}$) 
is consistent with periodic boundaries
at $-L$ and $L$.
The displacement $\overline{du}_{1}$ in~(\ref{eq:boundary_all}$_{1}$)
will be half of the shearing displacement that occurs 
across the full height $[-L,L]$ between the two
strains $\gamma_{\mathrm{o}}$ and $\gamma_{\mathrm{o}} + \overline{d\gamma}$.
The values of $G^{\mathrm{ep},-}$, 
$B_{2}^{\mathrm{ep},+}$, and $B_{2}^{\mathrm{ep},-}$
are the experimental values in Fig.~\ref{fig:G_load_unload} 
and Table~\ref{table:dgammapp_ep}.
The value of modulus $B_{1}^{\mathrm{ep}}$ was found to depend upon
the increments $d\gamma$ and $|d\Gammap|$, and although
the full nature of this relation could not be fully explored,
we used a form of $B_{1}^{\mathrm{ep}}$ that was consistent with
the data in Table~\ref{table:dgammapp_ep} and Fig.~\ref{fig:coupling_2}.
\par
The system of equations~(\ref{eq:dgammapp_ep}), 
(\ref{eq:equilibrium}), and~(\ref{eq:boundary_all}) admit
the trivial solution
\begin{equation}
du_{1}(x_{2}) = (x_{2}/L) \,\overline{du}_{1}\;,
\end{equation}
but the non-linear system may also have other solutions as well.
We used the Matlab package with its
Newton--Raphson iteration scheme to seek other solutions.
\par
The system~(\ref{eq:dgammapp_ep}), 
(\ref{eq:equilibrium}), and~(\ref{eq:boundary_all}) also
requires a choice of the loading modulus $G^{\mathrm{ep},+}$.
The Series~II experiments indicated that, when deformation is constrained,
the loading modulus is positive until the strain $\gamma$
reaches 2.7\% (Fig.~\ref{fig:G_load_unload}).
In unconstrained tests, non-persistent bands began to form
at a shearing strain of only 0.5\%, and a persistent
band was active at a strain of about 2\%.
We were unable, however, to find a stable band-like
solution to the boundary value 
problem
when the loading modulus was uniformly positive throughout the
region $[-L,L]$.
This result suggests two conditions that could at first initiate and then
later strengthen a shear band:
\begin{itemize}
\item
Local softening
could initiate a shear band.
The evidence of Section~\ref{sec:variation}
shows that local softening 
is a pervasive characteristic of granular materials
(Fig.~\ref{fig:scatter_shear}).
The results also show that after continued deformation,
regions of local weakening can latter harden and heal, freezing the shear band.
\item
Once initiated, a shear band could
become persistent if the local weakening is sustained
by progressively larger strains that would 
produce an irrecoverable softening of the material.
\end{itemize}
\par
\begin{figure*}
\centering\small
\includegraphics{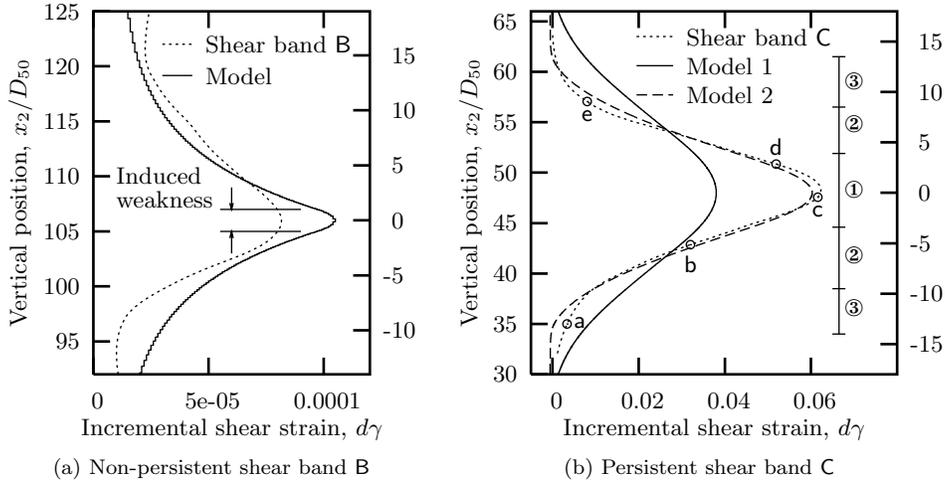}
\caption{Two one-dimensional models of a granular
material. (a)~Initiation of a non-persistent shear band.
The resulting strain profile is compared with the non-persistent
shear band~$\mathsf{B}$ of Fig.~\ref{fig:displacements_250}.  
(b)~Intensification of a persistent shear band between 
strains $\overline{\gamma}=2\%$--2.5\%.
The resulting strain profiles are compared with the shear band 
$\mathsf{C}$ of
Fig.~\ref{fig:displacements_250}.}
\label{fig:profiles_vs_theory}
\end{figure*}

The first possibility was investigated by 
assigning a positive value  of 3$p_{\mathrm{o}}$ to the loading
modulus $G^{\mathrm{ep},+}$ throughout the region $[0,L]$
except within a thin weak zone near the center of a possible 
band,
where we assign the slightly negative stiffness of
$G^{\mathrm{ep},+} = -1p_{\mathrm{o}}$ (Fig.~\ref{fig:profiles_vs_theory}a).
The positive value of 3$p_{\mathrm{o}}$ is the 
loading modulus that was measured in the constrained
Series~II tests at the strain
$\overline{\gamma} = 1.6\%$ (Fig.~\ref{fig:G_load_unload}a).
\par
A stable solution of the boundary value problem
is shown in Fig.~\ref{fig:profiles_vs_theory}a for
the conditions that were present at the appearance
of the non-persistent band~\textsf{B},
and this solution is roughly consistent with the measured shearing profile of
band~\textsf{B}.
These results show that a shear band can form in a region
of local weakness while the rest of the material
is undergoing a general hardening,
although the band may not be persistent.
\par
A second choice of the modulus $G^{\mathrm{ep},+}$
was used for investigating whether the same model
can predict the development of a persistent shear band.
We now consider the strain interval of
$\overline{d\gamma}=2.0$--2.5\%, which
includes both hardening and softening behaviors.
Shear band~$\mathsf{C}$ had formed and had become persistent
within this strain interval, and the
same range was considered in the previous 
Section~\ref{sec:gradients_in_band} and in 
Figs.~\ref{fig:phase_plane_band} and~\ref{fig:trajectory_200_250}.
The measured loading modulus $G^{\mathrm{ep},+}$ is shown
in Fig.~\ref{fig:G_load_unload}b
for strains greater than 2\%,
and these results suggest that a general softening occurs at strains of
2.7\% and beyond.
The dashed line in the figure gives the moduli
$G^{\mathrm{ep},+}$ that are used in the current model.
\par
Two solutions to the model are shown in Fig.~\ref{fig:profiles_vs_theory}b.
The solutions are compared with the strain profile that
was measured within the persistent shear band~$\mathsf{C}$
(Figs.~\ref{fig:displacements_250}, 
\ref{fig:phase_plane_band}, and~\ref{fig:trajectory_200_250}).
One solution uses the values of 
moduli~$B_{2}^{\mathrm{ep},+}$ and~$B_{2}^{\mathrm{ep},-}$
given in Table~\ref{table:dgammapp_ep}.
These ``Model~1'' values were independently measured in
the incremental sinusoid Series~IV tests,
but they produce a shear band that is somewhat thicker than shear
band~$\mathsf{C}$.
A second pair of moduli were inferred in Section~\ref{sec:gradients_in_band}
from the measurements of $d\gamma$ and $d\Gammapp$ within
band~$\mathsf{C}$ itself, and, not surprisingly, these values
produce a strain profile that more closely matches band~$\mathsf{C}$
(Model~2: $B_{2}^{\mathrm{ep},+}=-30D^{2}_{50}p_{\mathrm{o}}$,
$B_{2}^{\mathrm{ep},-}=-6D^{2}_{50}p_{\mathrm{o}}$).
\par
The regions~\ding{172}, \ding{173}, and~\ding{174} 
in Fig.~\ref{fig:profiles_vs_theory}b correspond to
those that were discussed in Section~\ref{sec:gradients_in_band}
(see Fig.~\ref{fig:trajectory_200_250}).
The material is softening within regions~\ding{172} and~\ding{173},
hardening within region~\ding{174}, and unloading
outside of the band.
The simple model demonstrates that a shear band can become
persistent when strains within the band are large enough to
bring the material into a state of sustained
softening, even while material near the edges and outside of the
band is still hardening or is unloading.
\section{Conclusions}  \label{sec:conclusions}
In the preceding sections we have used simulation experiments to 
explore the complex behavior of granular materials.
The experiments demonstrate several characteristics
of these materials.
\begin{enumerate}
\item
Strain localization is not an isolated phenomenon that only
accompanies failure.
At each stage of unconstrained loading, deformation
is concentrated within small regions.
Indeed, rather aggressive means are required to suppress
this tendency toward localization.
At low strains, intense shearing occurs within an
evolving network of microbands.
At moderate strains, the deformation is concentrated within much
thicker non-persistent shear bands.
At yet larger strains, shear bands become stationary and persistent.
\item
Granular materials are highly heterogeneous.
Local weakening, healing, and strengthening are neither infrequent nor
isolated occurrences, but are pervasive
characteristics of granular discontinua.
\item
Microbands and shear bands likely originate within weaker regions,
but these regions can later strengthen, perhaps arresting one
localization band while other bands are forming elsewhere.
\item
Granular materials are non-simple.
Shearing stress depends upon both the first and second gradients
of shear strain.
\item
The incremental influences of the shearing strain and its
first two gradients are all non-linear and are likely coupled.
\item
The influence of the first strain gradient suggests a lack of
central symmetry in the averaged local behavior.
\item
In spite of large fluctuations in particle rotations,
they conform, on average,
with the mean-field rotation, even in the presence of non-uniform
deformations.
\item
The average dilation that accompanies shearing is a function of the
local strain but is independent of the gradients of strain.
\item
An experimentally based one-dimensional continuum model
can capture certain features of localization
patterns.
A simple second-gradient linear model can predict the spacing
of microbands at small strains.
\item
A more complex non-linear model can predict the deformation profile of
a non-persistent shear band.
The model requires, however, the presence of localized softening
in the midst of general hardening.
\item
An even more complex model can produce the
deformation profile of a persistent shear band, within which 
large shearing strains produce a localized and unrecoverable
softening.
\end{enumerate}
\par
In short, the experiments show that granular materials are 
gradient-dependent, that this dependence is measurable,
and that the measurements can be applied to explain
characteristics of three different localization phenomena.
The experiments also reveal that the gradient-dependent behavior
is quite complex, with many observed features and non-linearities
that have not yet been
embodied in any current continuum, constitutive proposal.
The particle simulations could, of course, be expanded to
more general conditions, for example, by using three-dimensional
simulations of non-spherical particles.
Furthermore, the experiments have only probed the effect of the one-dimensional
gradients $u_{1,2}$, $u_{1,22}$, $u_{1,222}$, etc.,
and other gradients would need to be explored before volumetric
localization phenomena, such as compaction bands, could be modeled.
These and other important effects are left for future study.
%
%
%
%

\end{document}